\documentclass[prd, twocolumn, amsfonts, amsmath, amssymb, twoside, showpacs, nofootinbib]{revtex4}

\usepackage{bbm}
\usepackage{bm}
\usepackage{graphicx}
\usepackage{latexsym}
\usepackage{mathrsfs}

\newcommand{\MPIK}{Max-Planck-Institut f\"ur Kernphysik\\
                   Saupfercheckweg 1, 69117 Heidelberg, Germany\\ \ }

\newcommand{\3}{{\ss}}

\newcommand{\adjoint}[1]{{#1}^{\dagger}}

\newcommand{\C}{\mathcal{C}}

\newcommand{\ddd}[1]{d^3 #1}

\newcommand{\dddd}[1]{d^4 #1}

\newcommand{\dslash}{\not{\hbox{\kern-2pt $\partial$}}}

\renewcommand{\Im}{\mbox{Im}}

\newcommand{\intl}{\int\limits}

\newcommand{\KB}{Ka\-da\-noff-Baym\ }

\newcommand{\one}{\mathbbm{1}}

\newcommand{\PB}[2]{\left\{ #1 ;\; #2 \right\}_{PB}}

\newcommand{\Psibar}{{\bar{\Psi}}}

\renewcommand{\Re}{\mbox{Re}}

\newcommand{\sign}{\mbox{sign}}

\newcommand{\tr}{\mbox{tr}}

\begin{document}

\title{Comparison of Boltzmann Kinetics with Quantum Dynamics\\
for a Chiral Yukawa Model Far From Equilibrium}

\author{Manfred Lindner}
\email{lindner@mpi-hd.mpg.de}
\affiliation{\MPIK}

\author{Markus Michael M\"uller}
\email{Markus.Michael.Mueller@mpi-hd.mpg.de}
\affiliation{\MPIK}

\date{\today}

\begin{abstract}
Boltzmann equations are often used to describe the non-equilibrium
time-evolution of many-body systems in particle physics. Prominent examples
are the computation of the baryon asymmetry of the universe and the evolution
of the quark-gluon plasma after a relativistic heavy ion collision. However,
Boltzmann equations are only a classical approximation of the quantum
thermalization process, which is described by so-called Kadanoff-Baym
equations. This raises the question how reliable Boltzmann equations are as
approximations to the complete Kadanoff-Baym equations. Therefore, we present
in this article a detailed comparison of Boltzmann and Kadanoff-Baym equations
in the framework of a chirally invariant Yukawa-type quantum field theory
including fermions and scalars. The obtained numerical results reveal
significant differences between both types of equations. Apart from
quantitative differences, on a qualitative level the late-time universality
respected by Kadanoff-Baym equations is severely restricted in the case of
Boltzmann equations. Furthermore, Kadanoff-Baym equations strongly separate
the time scales between kinetic and chemical equilibration. In contrast to
this standard Boltzmann equations cannot describe the process of
quantum-chemical equilibration, and consequently also cannot feature the above
separation of time scales.
\end{abstract}

\maketitle

\section{Introduction}

This article is an extension of our previous studies~\cite{Lindner:2005kv},
where we performed a detailed comparison of Boltzmann and \KB equations in the
framework of a real scalar $\Phi^4$ quantum field theory. The motivation for
these studies is a better understanding of processes like leptogenesis or
preheating in the early universe~\cite{Sakharov:1967dj,Fukugita:1986hr,
Buchmuller:2000nd,Berges:2002cz}, or the evolution of the quark-gluon plasma
after relativistic heavy-ion collisions~\cite{Arsene:2004fa,Back:2004je,
Adams:2005dq,Adcox:2004mh}. All these phenomena require the description of
many-particle systems out of thermal equilibrium. The standard means to deal
with this non-equilibrium situation are Boltzmann equations. However, it is
well known that (classical) Boltzmann equations suffer from several
shortcomings as compared to their quantum mechanical generalizations,
so-called \KB equations. This motivates a comparison of Boltzmann and \KB
equations in order to assess the reliability of quantitative predictions based
on standard Boltzmann techniques. In the present work we generalize our
previous results to the case of a chirally invariant Yukawa-type quantum field
theory coupling fermions with scalars. More precisely, we consider a globally
$SU(2) \times SU(2) \times U(1)$ symmetric quantum field theory, which offers
two interpretations for the particle content: On one hand, in the context of
leptogenesis one might think of a single generation of leptons and a Higgs
bidoublet~\cite{Deshpande:1990ip}. On the other hand, this theory is
equivalent to the linear $\sigma$-model~\cite{Schwinger:1957em,
polkinghorne1958a,Gell-Mann:1960np}, which can be used to describe low-energy
quark-meson dynamics in two-flavor QCD. In any case, this work can be regarded
as a further step to approach more realistic theories, which can be used to
describe the phenomena motivating our studies.

What are the shortcomings of Boltzmann equations? Originally, Boltzmann
equations have been designed for the description of the non-equilibrium
time-evolution of dilute gases of classical particles. As such, their range
of validity must be scrutinized once quantum effects become relevant. This is
certainly the case for elementary particles playing the central role in
phenomena like leptogenesis or the quark-gluon plasma. As already indicated
above, the quantum dynamics of such systems is described by so-called \KB
equations. Employing a sequence of approximations, Boltzmann equations can be
derived from \KB equations~\cite{baymKadanoff1962a,Danielewicz:1982kk,
Ivanov:1999tj,Knoll:2001jx, Blaizot:2001nr}. However, it is important to note
that these approximations might be neither justifiable nor controllable, and
sometimes even inconsistent. After all, standard Boltzmann equations take only
on-shell processes into account, feature spurious constants of motion, and
introduce irreversibility by implying the assumption of molecular chaos 
(``Sto\3\-zahl\-an\-satz'')~\cite{landauLifshitzVolume10,balescu1975a,
deGroot1980a,kreuzer1981a}. In contrast to this, \KB equations are
time-reversal invariant and take memory and off-shell effects  into
account~\cite{Berges:2000ur,kohler1995a,kohler1996a,Aarts:2001qa}. Therefore,
one should perform a detailed comparison of Boltzmann and \KB 
equations~\cite{Danielewicz:1982ca,kohler1995a,kohler1996a,Morawetz:1998em,
Juchem:2003bi,Lindner:2005kv,Muller:2006ny}.

Due to the complexity of the problem, in a first step we restricted ourselves
to a real scalar $\Phi^4$ quantum field theory in $3+1$ space-time
dimensions~\cite{Lindner:2005kv}. Of course, in this framework one can neither
describe the phenomenon of leptogenesis nor thermalization of the quark-gluon
plasma after a relativistic heavy ion collision. Nevertheless, it certainly
allowed to perform a detailed comparison of Boltzmann and \KB equations, which
revealed interesting phenomena to be investigated in more realistic theories.
We found considerable differences in the results furnished by the Boltzmann
and \KB equations: On a quantitative level, we found that the Boltzmann
equation predicts significantly larger thermalization times than the
corresponding \KB equations. On a qualitative level we could verify that \KB
equations respect full late-time universality~\cite{Berges:2000ur,
Berges:2004yj} and strongly separate the time scales between kinetic and
chemical equilibration~\cite{Berges:2004ce}. In the case of a real scalar
$\Phi^4$ quantum field theory the Boltzmann equation includes only two-particle
scattering processes, which conserve the total particle number. This spurious
constant of motion severely constrains the evolution of the particle number
distribution. As a result, the Boltzmann equation respects only a restricted
universality, fails to describe the process of quantum-chemical equilibration,
and does not separate any time scales.

\begin{figure}
  \includegraphics[width=7cm]{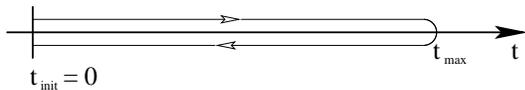}
  \caption{\label{fig3.1}Closed real-time path $\C$. This time path was 
  invented by Schwinger~\cite{Schwinger:1960qe} and applied to 
  non-equilibrium problems by Keldysh~\cite{Keldysh:1964ud}. In order to 
  avoid the doubling of the degrees of freedom, we use the form presented 
  in Ref.~\cite{Danielewicz:1982kk}.}
\end{figure} 
In the present work we extent our comparison of Boltzmann and \KB equations
to a chirally invariant Yukawa-type quantum field theory coupling fermions
with scalars. We start from the 2PI effective action~\cite{Jackiw:1974cv,
Cornwall:1974vz} and derive the \KB equations by requiring that the 2PI
effective action be stationary with respect to variations of the complete
connected two-point functions~\cite{Calzetta:1986cq,Berges:2002wr}.
First, this guarantees that the \KB equations conserve the average energy
density as well as global charges~\cite{baym1961a,Baym:1962sx,Ivanov:1998nv},
and second, the 2PI effective action has proven to be an efficient and
reliable tool for the description of quantum fields out of thermal equilibrium
in numerous previous treatments~\cite{Berges:2000ur,Berges:2001fi,
Berges:2002wr,Aarts:2001yn,Aarts:2003bk}. In order to derive the corresponding
Boltzmann equations, subsequently one has to employ a first-order gradient
expansion, a Wigner transformation, the \KB ansatz and the quasi-particle 
approximation~\cite{baymKadanoff1962a,Danielewicz:1982kk,Ivanov:1999tj,
Knoll:2001jx,Blaizot:2001nr}. While Boltzmann equations describe the time 
evolution of particle number distributions, \KB equations describe 
the evolution of the complete quantum mechanical two-point functions of the 
system. However, one can define effective particle number distributions, which
can be obtained from the complete propagators and their time derivatives
evaluated at equal times~\cite{Berges:2001fi,Berges:2002wr}. Finally, we solve
the Boltzmann and the \KB equations numerically for highly symmetric systems
in 3+1 space-time dimensions and compare their predictions on the evolution of
these systems for various initial conditions.

\section{2PI Effective Action}

We consider a globally $SU(2)_L \times SU(2)_R \times U(1)_{B-L}$ symmetric 
quantum field theory with one generation of chiral leptons and a Higgs 
bidoublet with Dirac-Yukawa interactions~\cite{Deshpande:1990ip}. The Dirac
fields are denoted with $\Psi_l^{\alpha} \left( x \right)$, where $\alpha$ is
a Dirac index and $l \in \left\{ \nu, e \right\}$ denotes the type of the
leptons. Using the Pauli matrices, the (complex) Higgs bidoublet can be
parameterized by real scalar fields denoted with $\Phi_a \left( x \right)$,
where $a \in \left\{ 0, \ldots, 3 \right\}$. In this notation the Lagrangian
density takes the form\footnote{The form of the kinetic term indicates that
we use the Minkowski metric where the time-time component is negative.
$\sigma_1$, $\sigma_2$ and $\sigma_3$ are the usual Pauli matrices, while
$\sigma_0 = i \one$. In addition to the Dirac $\gamma$ matrices we will
frequently use $\beta = i \gamma^0$, and
$P_L = \frac{1}{2} \left( \one + \gamma_5 \right)$ and
$P_R = \frac{1}{2} \left( \one - \gamma_5 \right)$.}
\begin{eqnarray*}
        \mathscr{L}
  & = & - \Psibar \dslash \Psi - \frac{1}{2} \Big( \partial_{\mu} \Phi_a \Big) \Big( \partial^{\mu} \Phi_a \Big) - \frac{1}{2} m_B^2 \Phi_a \Phi_a \\
  &   & {} - \lambda \left( \Phi_a \Phi_a \right)^2 - i \eta \Psibar \Phi_a \left( \sigma_a P_R - \adjoint{\sigma}_a P_L \right) \Psi \;.
\end{eqnarray*}
Although we refer to the scalar fields as Higgs fields, we would like to note
that this theory is equivalent to the linear 
$\sigma$-model~\cite{Schwinger:1957em,polkinghorne1958a,Gell-Mann:1960np},
which can be used to describe low-energy quark-meson dynamics in two-flavor
QCD. This and a similar model have been considered in a related context in 
Refs.~\cite{Berges:2002wr,Berges:2004ce}.

\begin{figure}
  \centering
  \includegraphics{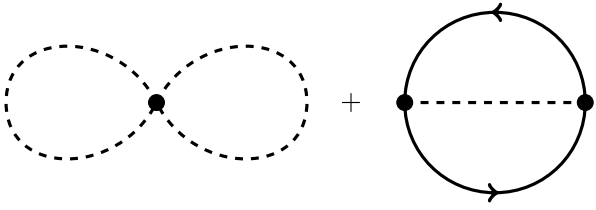}
  \caption{\label{fig3.2}Two-loop contribution to $\Gamma_2 \left[ G, S 
  \right]$. Full (dashed) lines represent the complete connected lepton 
  (Higgs) propagator $S$ ($G$).}
\end{figure}

Since we will compute the evolution of the two-point Green's functions for 
non-equilibrium initial conditions, already the classical action has to be
defined on the closed Schwinger-Keldysh real-time contour, shown in 
Figure~\ref{fig3.1}. The free inverse propagators can then be read off the free 
part of the classical action:
\begin{eqnarray*}
        I_0
  & = & - \intl_{\C} \dddd{x} \; \dddd{y} \Bigg[ \Psibar_l \left( x \right) S_{0, lm}^{-1} \left( x, y \right) \Psi_m \left( y \right) \\
  &   & {} + \frac{1}{2} \Phi_a \left( x \right) G_{0, ab}^{-1} \left( x, y \right) \Phi_b \left( y \right) \Bigg] \;,
\end{eqnarray*}
where the inverse free propagators are given by
\begin{equation} \label{eq3.2}
  G_{0, ab}^{-1} \left( x, y \right) = \left( \partial_{x^{\mu}} \partial_{y_{\mu}} + m_B^2 \right) \delta_{\C} \left( x - y \right) \delta_{ab}
\end{equation}
and
\begin{equation} \label{eq3.1}
  S_{0, lm}^{-1} \left( x, y \right) = \dslash_x \delta_{\C} \left( x - y \right) \delta_{lm} \;.
\end{equation}
We consider a system without symmetry breaking, i.~e.\ $\left\langle \Phi_a
\left( x \right) \right\rangle = 0$. In this case the full connected
Schwinger-Keldysh propagators are given by
\begin{equation} \label{eq3.23}
  G_{ab} \left( x, y \right) = \left\langle T_{\C} \left\{ \Phi_a \left( x \right) \Phi_b \left( y \right) \right\} \right\rangle
\end{equation}
and
\begin{equation} \label{eq3.22}
  S_{lm}^{\alpha \beta} \left( x, y \right) = \left\langle T_{\C} \left\{ \Psi_l^{\alpha} \left( x \right) \Psibar_m^{\beta} \left( y \right) \right\} \right\rangle \;,
\end{equation}
so that the 2PI effective action can be written in the form
\begin{eqnarray}
        \Gamma \left[ G, S \right]
  & = & \frac{i}{2} \tr_{\C} \log_{\C} \left[ G^{-1} \right] - \frac{1}{2} \tr_{\C} \left[ G_0^{-1} G \right] \nonumber \\
  &   & {} - i \tr_{\C} \log_{\C} \left[ S^{-1} \right] + \tr_{\C} \left[ S_0^{-1} S \right] \nonumber \\
  &   & {} + \Gamma_2 \left[ G, S \right] + const \;. \label{eq3.26}
\end{eqnarray}
The square brackets indicate that the trace, the logarithm and the product of
the propagators have to be taken in the functional sense, and the subscript
$\C$ reminds us that integrals over time are running along the closed 
real-time contour. $i \Gamma_2 \left[ G, S \right]$ is the sum of all 
two-particle irreducible vacuum diagrams, where internal lines represent the 
complete connected propagators $S$ and $G$. In this work we apply the loop
expansion of the 2PI effective action up to two-loop order. The diagrams 
contributing to $\Gamma_2$ in this approximation are shown in 
Figure~\ref{fig3.2}. Using the abbreviation
\[ H_{a,lm}^{\alpha \beta} = i \eta \left[ \left( \sigma_a \right)_{lm} P_R^{\alpha \beta} - \left( \adjoint{\sigma}_a \right)_{lm} P_L^{\alpha \beta} \right] \]
we find
\begin{eqnarray*}
  \lefteqn{\Gamma_2 \left[ G, S \right] = - \lambda \intl_{\C} \dddd{x} \Big[ G_{aa} \left( x, x \right) G_{bb} \left( x, x \right)} \\
  &   & {} + 2 G_{ab} \left( x, x \right) G_{ab} \left( x, x \right) \Big] \\
  &   & {} - \frac{i}{2} \intl_{\C} \dddd{x} \; \dddd{y} \; G_{ab} \left( x, y \right) \tr \Big( H_a S \left( x, y \right) H_b S \left( y, x \right) \Big) \;,
\end{eqnarray*}
where the trace runs over Dirac and lepton type indices.

\section{\KB Equations}

\begin{figure}
  \centering
  \includegraphics[width=75mm]{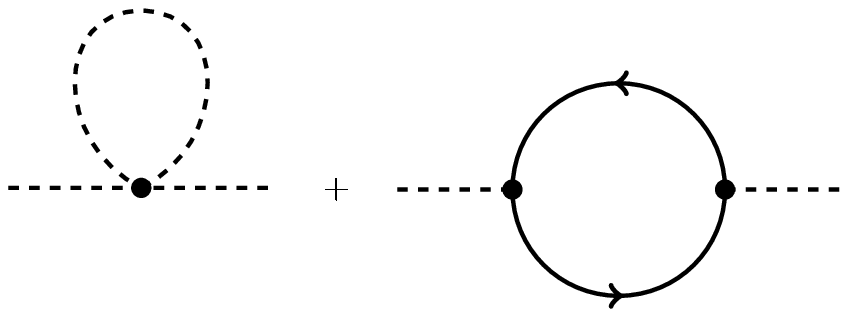}
  \caption{\label{fig3.3}One-loop contribution to the Higgs self-energy.}
  \vspace{10mm}
  \includegraphics{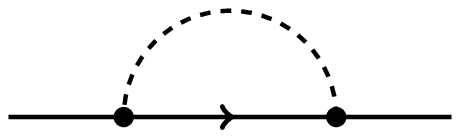}
  \caption{\label{fig3.4}One-loop contribution to the lepton self-energy.}
\end{figure}
The equations of motion for the complete propagators read
\begin{equation} \label{eq3.28}
  \frac{\delta \Gamma \left[ G, S \right]}{\delta G_{ba} \left( y, x \right)} = 0 \quad \mbox{and} \quad \frac{\delta \Gamma \left[ G, S \right]}{\delta S_{ml}^{\beta \alpha} \left( y, x \right)} = 0 \;.
\end{equation}
They are equivalent to the corresponding self-consistent Schwinger-Dyson 
equations
\begin{equation} \label{eq3.21}
  G^{-1}_{ab} \left( x, y \right) = i G^{-1}_{0,ab} \left( x, y \right) - \Pi_{ab} \left( x, y \right)
\end{equation}
and
\begin{equation} \label{eq3.15}
  S^{-1}_{lm} \left( x, y \right) = i S^{-1}_{0,lm} \left( x, y \right) - \Sigma_{lm} \left( x, y \right) \;,
\end{equation}
where the proper self-energies are given by
\begin{eqnarray} 
  \lefteqn{\Pi_{ab} \left( x, y \right) = 2 i \frac{\delta \Gamma_2 \left[ G, S \right]}{\delta G_{ba} \left( y, x \right)}} \; \nonumber \\
  & = & - 4 \lambda i \delta_{\C} \left( x - y \right) \Big( G_{dd} \left( x, x \right) \delta_{ab} + 2 G_{ab} \left( x, x \right) \Big) \nonumber \\
  &   & {} + \tr \Big( H_a S \left( x, y \right) H_b S \left( y, x \right) \Big) \label{eq3.13}
\end{eqnarray}
and
\begin{eqnarray}
  \lefteqn{\Sigma_{lm}^{\alpha \beta} \left( x, y \right) = - i \frac{\delta \Gamma_2 \left[ G, S \right]}{\delta S_{ml}^{\beta \alpha} \left( y, x \right)}} \; \nonumber \\
  & = & - H_{a, lk}^{\alpha \gamma} H_{b, nm}^{\delta \beta} S_{kn}^{\gamma \delta} \left( x, y \right) G_{ab} \left( x, y \right) \;. \label{eq3.14}
\end{eqnarray}
Next, we define the spectral function\footnote{From the definition of the 
Higgs spectral-function we see that it is antisymmetric in the sense that 
$G_{\varrho, ba} \left( y, x \right) = - G_{\varrho, ab} \left( x, y \right)$.
Furthermore, the canonical equal-time commutation relations give $\left( 
G_{\varrho, ab} \left( x, y \right) \right)_{x^0 = y^0} = 0$ and $\left( 
\partial_{y^0} G_{\varrho, ab} \left( x, y \right) \right)_{x^0 = y^0} = - 
\delta_{ab} \delta^3 \left( \bm{x} - \bm{y} \right)$.}
\[ G_{\varrho, ab} \left( x, y \right) = i \left\langle \left[ \Phi_a \left( x \right), \Phi_b \left( y \right) \right]_- \right\rangle \]
and the statistical propagator\footnote{In contrast to the spectral function,
the statistical Higgs-pro\-pa\-ga\-tor is symmetric in the sense that $G_{F, ba} 
\left( y, x \right) = G_{F, ab} \left( x, y \right)$.}
\[ G_{F, ab} \left( x, y \right) = \frac{1}{2} \left\langle \left[ \Phi_a \left( x \right), \Phi_b \left( y \right) \right]_+ \right\rangle \]
for the Higgs bosons, so that we can write the complete Higgs propagator as
\begin{equation} \label{eq3.20}
  G \left( x, y \right) = G_F \left( x, y \right) - \frac{i}{2} \sign_{\C} \left( x^0 - y^0 \right) G_{\varrho} \left( x, y \right) \;.
\end{equation}
In the case of real scalar fields the spectral function and the statistical 
propagator are real-valued quantities~\cite{Berges:2001fi}.
In a similar way, we also define the spectral function\footnote{The 
adjoint of the lepton spectral-function is given by $\adjoint{S}_{\varrho, 
lm} \left( x, y \right) = - \beta S_{\varrho, ml} \left( y, x \right) \beta$. 
Furthermore, the canonical equal-time anti-commutation relations give
$\left( S_{\varrho, lm} \left( x, y \right) \right)_{x^0 = y^0} = i \beta
\delta \left( \bm{x} - \bm{y} \right) \delta_{lm}$.}
\begin{equation} \label{eq3.16}
  S_{\varrho, lm}^{\alpha \beta} \left( x, y \right) = i \left\langle \left[ \Psi_l^{\alpha} \left( x \right), \Psibar_m^{\beta} \left( y \right) \right]_+ \right\rangle
\end{equation}
and the statistical propagator\footnote{The adjoint of the statistical 
lepton-propagator is given by $\adjoint{S}_{F, lm} \left( x, y \right) = 
\beta S_{F, ml} \left( y, x \right) \beta$.}
\begin{equation} \label{eq3.17}
  S_{F, lm}^{\alpha \beta} \left( x, y \right) = \frac{1}{2} \left\langle \left[ \Psi_l^{\alpha} \left( x \right), \Psibar_m^{\beta} \left( y \right) \right]_- \right\rangle
\end{equation}
for the leptons, so that we can decompose the complete lepton propagator 
according to
\begin{equation} \label{eq3.18}
  S \left( x, y \right) = S_F \left( x, y \right) - \frac{i}{2} \sign_{\C} \left( x^0 - y^0 \right) S_{\varrho} \left( x, y \right) \;.
\end{equation}
Then, using Eqs.~(\ref{eq3.20}) and (\ref{eq3.18}), we can decompose the Higgs
self-energy (\ref{eq3.13}) as well as the lepton self-energy (\ref{eq3.14})
according to:
\begin{eqnarray*}
  \Pi \left( x, y \right) 
  & = & - i \delta_{\C} \left( x - y \right) \Pi^{(local)} \left( x \right) \\
  &   & {} + \Pi_F \left( x, y \right) - \frac{i}{2} \sign_{\C} \left( x^0 - y^0 \right) \Pi_{\varrho} \left( x, y \right)
\end{eqnarray*}
and
\[ \Sigma \left( x, y \right) = \Sigma_F \left( x, y \right) - \frac{i}{2} \sign_{\C} \left( x^0 - y^0 \right) \Sigma_{\varrho} \left( x, y \right) \;. \]
The local part of the Higgs self-energy causes a mass shift only, wherefore we
define the effective mass by
\begin{eqnarray}
  \lefteqn{M_{ab}^2 \left( x \right) = m_B^2 \delta_{ab} + \Pi_{ab}^{(local)} \left( x \right)} \; \label{eq3.12} \\
  & = & m_B^2 \delta_{ab} + 4 \lambda \Big( G_{dd} \left( x, x \right) \delta_{ab} + 2 G_{ab} \left( x, x \right) \Big) \;. \nonumber
\end{eqnarray}
After convoluting Eqs.~(\ref{eq3.21}) and (\ref{eq3.15}) from the right with 
the corresponding complete propagators, we observe that both equations split into 
two complementary evolution equations for the statistical propagators and the 
spectral functions, respectively~\cite{Berges:2002wr}:
\begin{eqnarray} 
  \lefteqn{\left( - \partial_{x^\mu} \partial_{x_\mu} \delta_{ac} + M_{ac}^2 \left( x \right) \right) G_{F, cb} \left( x, y \right)} \nonumber \\
  & = & \intl_{0}^{y^0} \dddd{z} \; \Pi_{F, ac} \left( x, z \right) G_{\varrho, cb} \left( z, y \right) \nonumber \\
  &   & {} - \intl_{0}^{x^0} \dddd{z} \; \Pi_{\varrho, ac} \left( x, z \right) G_{F, cb} \left( z, y \right) \label{eq3.6} \;,
\end{eqnarray}
\begin{eqnarray}
  \lefteqn{\left( - \partial_{x^\mu} \partial_{x_\mu} \delta_{ac} + M_{ac}^2 \left( x \right) \right) G_{\varrho, cb} \left( x, y \right)} \nonumber \\
  & = & - \intl_{y^0}^{x^0} \dddd{z} \; \Pi_{\varrho, ac} \left( x, z \right) G_{\varrho, cb} \left( z, y \right) \;, \label{eq3.7}
\end{eqnarray}
\begin{eqnarray}
  \lefteqn{\dslash_x S_{F, lm} \left( x, y \right) = \intl_0^{y^0} \dddd{z} \; \Sigma_{F, lk} \left( x, z \right) S_{\varrho, km} \left( z, y \right)} \nonumber \\
  &   & {} - \intl_0^{x^0} \dddd{z} \; \Sigma_{\varrho, lk} \left( x, z \right) S_{F, km} \left( z, y \right) \hspace{15mm} \label{eq3.4}
\end{eqnarray}
and
\begin{equation} \label{eq3.5}
  \dslash_x S_{\varrho, lm} \left( x, y \right) = - \intl_{y^0}^{x^0} \dddd{z} \; \Sigma_{\varrho, lk} \left( x, z \right) S_{\varrho, km} \left( z, y \right) \;.
\end{equation}
Nowadays, it is practically impossible to solve the \KB equations numerically
in this general form. However, for initial conditions which are invariant
under spatial translations, spatial rotations, parity, charge conjugation, and
chiral transformations, the propagators take the form~\cite{Berges:2002wr}
\begin{equation} \label{eq3.44}
  G_{ab} \left( x, y \right) = \int \frac{\ddd{k}}{\left( 2 \pi \right)^3} \; \exp \Big( i \bm{k} \left( \bm{x} - \bm{y} \right) \Big) G \left( x^0, y^0, k \right) \delta_{ab}
\end{equation}
and
\begin{eqnarray}
  \lefteqn{S_{lm} \left( x, y \right) = \int \frac{\ddd{k}}{\left( 2 \pi \right)^3} \; \exp \Big( i \bm{k} \left( \bm{x} - \bm{y} \right) \Big)} \; \label{eq3.45} \\
  & & {} \times \left( S_V^0 \left( x^0, y^0, k \right) \gamma_0 - i \frac{k^j}{k} S_V \left( x^0, y^0, k \right) \gamma_j \right) \; \delta_{lm} \;, \nonumber
\end{eqnarray}
where $k = \left| \bm{k} \right|$. The index $V$ indicates that $S_V^{\mu}$ 
would transform as a vector under a Lorentz transformation. Due to CP 
invariance, the statistical and spectral vector components of the lepton 
propagator satisfy
\begin{equation} \label{eq3.46}
  \begin{array}{rcrcr}
    S_{V, F}^0 \left( x^0, y^0, k \right)       \! & = & \! - S_{V, F}^0 \left( y^0, x^0, k \right)     \! & = & \! S_{V, F}^{0^{\scriptstyle \ast}} \left( x^0, y^0, k \right) \\[2mm]
    S_{V, F} \left( x^0, y^0, k \right)         \! & = & \! S_{V, F} \left( y^0, x^0, k \right)         \! & = & \! S_{V, F}^{\ast} \left( x^0, y^0, k \right) \\[2mm]
    S_{V, \varrho}^0 \left( x^0, y^0, k \right) \! & = & \! S_{V, \varrho}^0 \left( y^0, x^0, k \right) \! & = & \! S_{V, \varrho}^{0^{\scriptstyle \ast}} \left( x^0, y^0, k \right) \\[2mm]
    S_{V, \varrho} \left( x^0, y^0, k \right)   \! & = & \! - S_{V, \varrho} \left( y^0, x^0, k \right) \! & = & \! S_{V, \varrho}^{\ast} \left( x^0, y^0, k \right)
  \end{array}
\end{equation}
Thus, the explicit factor of $i$ makes $S_{V, F}^0$, $S_{V, F}$, 
$S_{V, \varrho}^0$ and $S_{V, \varrho}$ real-valued quantities:
\begin{equation} \label{eq3.50}
  \Im \left( S_{V, F}^0 \right) = \Im \left( S_{V, F} \right) = \Im \left( S_{V, \varrho}^0 \right) = \Im \left( S_{V, \varrho} \right) = 0 \;.
\end{equation}
Furthermore, the canonical equal-time anti-commutation relations for fermion
fields imply
\begin{equation} \label{eq3.37}
  S_{V, \varrho}^0 \left( t, t, k \right) = 1 \qquad \mbox{and} \qquad S_{V, \varrho} \left( t, t, k \right) = 0 \;,
\end{equation}
the second equality being consistent with the anti-symmetry of the space-like
vector-component of the lepton spectral function, cf.~Eq.~(\ref{eq3.46}). Of
course, the relations (\ref{eq3.44}), (\ref{eq3.45}), (\ref{eq3.46}) and
(\ref{eq3.50}) also hold for the corresponding self energies, so that the 
\KB equations can be simplified drastically. The simplified 
\KB equations for the Higgs propagator read~\cite{Berges:2002wr}
\begin{widetext}
\begin{eqnarray} 
        \lefteqn{\left[ \partial^2_{x^0} + k^2 + M^2 \left( x^0 \right) \right] G_F \left( x^0, y^0, k \right)} \nonumber \\
  & = & \intl_{0}^{y^0} dz^0 \; \Pi_F \left( x^0, z^0, k \right) G_{\varrho} \left( z^0, y^0, k \right) - \intl_{0}^{x^0} dz^0 \; \Pi_{\varrho} \left( x^0, z^0, k \right) G_F \left( z^0, y^0, k \right) \label{eq3.38}
\end{eqnarray}
\\ and \\
\begin{equation} \label{eq3.39}
  \left[ \partial^2_{x^0} + k^2 + M^2 \left( x^0 \right) \right] G_{\varrho} \left( x^0, y^0, k \right) = - \intl_{y^0}^{x^0} dz^0 \; \Pi_{\varrho} \left( x^0, z^0, k \right) G_{\varrho} \left( z^0, y^0, k \right) \;.
\end{equation}
In the same way, the 128 complex-valued \KB equations (\ref{eq3.4}) and
(\ref{eq3.5}) for the lepton propagator can be reduced to the following 4
real-valued equations~\cite{Berges:2002wr}:
\vspace{2mm}
\begin{eqnarray}
        \lefteqn{\partial_{x^0} S_{V, F}^0 \left( x^0, y^0, k \right) + k S_{V, F} \left( x^0, y^0, k \right)} \; \label{eq3.40} \\
  & = & {} - \intl_0^{y^0} dz^0 \; \Big[ \Sigma_{V, F}^0 \left( x^0, z^0, k \right) S_{V, \varrho}^0 \left( z^0, y^0, k \right) + \Sigma_{V, F} \left( x^0, z^0, k \right) S_{V, \varrho} \left( z^0, y^0, k \right) \Big] \nonumber \\
  &   & {} + \intl_0^{x^0} dz^0 \; \Big[ \Sigma_{V, \varrho}^0 \left( x^0, z^0, k \right) S_{V, F}^0 \left( z^0, y^0, k \right) + \Sigma_{V, \varrho} \left( x^0, z^0, k \right) S_{V, F} \left( z^0, y^0, k \right) \Big] \nonumber \;,
\end{eqnarray}
\\
\begin{eqnarray}
        \lefteqn{\partial_{x^0} S_{V, F} \left( x^0, y^0, k \right) - k S_{V, F}^0 \left( x^0, y^0, k \right)} \; \label{eq3.41} \\
  & = & \intl_0^{y^0} dz^0 \; \Big[ \Sigma_{V, F} \left( x^0, z^0, k \right) S_{V, \varrho}^0 \left( z^0, y^0, k \right) - \Sigma_{V, F}^0 \left( x^0, z^0, k \right) S_{V, \varrho} \left( z^0, y^0, k \right) \Big] \nonumber \\
  &   & {} - \intl_0^{x^0} dz^0 \; \Big[ \Sigma_{V, \varrho} \left( x^0, z^0, k \right) S_{V, F}^0 \left( z^0, y^0, k \right) - \Sigma_{V, \varrho}^0 \left( x^0, z^0, k \right) S_{V, F} \left( z^0, y^0, k \right) \Big] \nonumber \;,
\end{eqnarray}
\\
\begin{eqnarray}
        \lefteqn{\partial_{x^0} S_{V, \varrho}^0 \left( x^0, y^0, k \right) + k S_{V, \varrho} \left( x^0, y^0, k \right)} \; \label{eq3.42} \\
  & = & \intl_{y^0}^{x^0} dz^0 \; \Big[ \Sigma_{V, \varrho}^0 \left( x^0, z^0, k \right) S_{V, \varrho}^0 \left( z^0, y^0, k \right) + \Sigma_{V, \varrho} \left( x^0, z^0, k \right) S_{V, \varrho} \left( z^0, y^0, k \right) \Big] \nonumber
\end{eqnarray}
\\ and \\
\begin{eqnarray}
        \lefteqn{\partial_{x^0} S_{V, \varrho} \left( x^0, y^0, k \right) - k S_{V, \varrho}^0 \left( x^0, y^0, k \right)} \; \label{eq3.43} \\
  & = & - \intl_{y^0}^{x^0} dz^0 \; \Big[ \Sigma_{V, \varrho} \left( x^0, z^0, k \right) S_{V, \varrho}^0 \left( z^0, y^0, k \right) - \Sigma_{V, \varrho}^0 \left( x^0, z^0, k \right) S_{V, \varrho} \left( z^0, y^0, k \right) \Big] \nonumber \;.
\end{eqnarray}
\end{widetext}
The expressions for the Higgs and lepton self-energies are given in
App.~\ref{sect3.2}. As explained in more detail in Ref.~\cite{Berges:2002wr},
one can define an effective kinetic energy distribution $\omega \left( t, k
\right)$, as well as effective scalar and fermion particle number
distributions $n_s \left( t, k \right)$ and $n_f \left( t, k \right)$, which
can be obtained from the statistical propagators according to
\begin{equation} \label{eq3.47}
  \omega^2 \left( t, k \right) = \left( \frac{\partial_{x^0} \partial_{y^0} G_F \left( x^0, y^0, k \right)}{G_F \left( x^0, y^0, k \right)} \right)_{x^0 = y^0 = t} \;,
\end{equation}
\begin{equation} \label{eq3.48}
  n_s \left( t, k \right) = \omega \left( t, k \right) G_F \left( t, t, k \right) - \frac{1}{2} \;.
\end{equation}
and
\begin{equation} \label{eq3.49}
  n_f \left( t, k \right) = \frac{1}{2} - S_{V, F} \left( t, t, k \right) \;.
\end{equation}
The definition of such particle numbers is necessary in order to make contact
to Boltzmann equations, e.~g.~when comparing numerical solutions of Boltzmann
and \KB equations, which we will do in Sect.~\ref{sect3.1}. We emphasize,
however, that the \KB equations are self-consistent evolution equations for
the complete propagators of our system, and that one has to follow the
evolution of the two-point functions throughout the complete $x^0$-$y^0$-plane
(of course, constrained to the part with $x^0 \ge 0$ and $y^0 \ge 0$). One can
then follow the evolution of the effective particle number densities along the
bisecting line of this plane.

\section{Boltzmann Equations \label{sect3.5}}

In this section we briefly sketch the standard way of deriving Boltzmann 
equations from \KB equations~\cite{baymKadanoff1962a,Danielewicz:1982kk,
Blaizot:2001nr,Berges:2004pu,Prokopec:2003pj,Prokopec:2004ic}. One has to
employ a Wigner transformation, a first-order gradient expansion, the \KB
ansatz and the quasi-particle approximation.

First, we subtract Eq.~(\ref{eq3.38}) (Eq.~(\ref{eq3.40})) with $x^0$ and 
$y^0$ interchanged from Eq.~(\ref{eq3.38}) (Eq.~(\ref{eq3.40})). Then we
re-parameterize the propagators and the self energies by center and relative
times, e.~g.
\[ G \left( u^0, v^0, k \right) = G \left( \frac{u^0 + v^0}{2}, u^0 - v^0, k \right) \;. \]
Next, we define the center time $t = \frac{x^0 + y^0}{2}$ and the relative 
time $s^0 = x^0 - y^0$, and observe on the left hand side of the difference 
equations that
\[ \partial_{x^0} \partial_{x^0} - \partial_{y^0} \partial_{y^0} = 2 \partial_t \partial_{s^0} \]
and 
\[ \partial_{x^0} + \partial_{y^0} = \partial_t \]
are automatically of first order in $\partial_t$. Furthermore, we Taylor 
expand the effective masses on the left hand side of the difference equation
for the scalars as well as the propagators and self energies on the right hand
sides of both difference equations to first order in $\partial_t$ around $t$. 
After that, we Fourier transform the difference equations with respect to the 
relative time $s^0$. The Wigner transformed scalar statistical propagator and 
scalar spectral function are given by
\begin{eqnarray*}
        G_F \left( t, \omega, k \right) 
  & = & \int ds^0 \; \exp \left( i \omega s^0 \right) G_F \left( t, s^0, k \right) \;, \\
        G_{\varrho} \left( t, \omega, k \right) 
  & = & - i \int ds^0 \; \exp \left( i \omega s^0 \right) G_{\varrho} \left( t, s^0, k \right) \;.
\end{eqnarray*}
As $G_{\varrho} \left( t, s^0, k \right)$ is a real-valued odd function of 
the relative time $s^0$, we introduced an explicit factor of $-i$ in order to 
make its Wigner transform again a real-valued function. For similar reasons we 
also introduce a factor of $-i$ for the Wigner transforms of $S_{V, \varrho} 
\left( t, s^0, k \right)$, $S_{V, F}^0 \left( t, s^0, k \right)$, $S_{V, R}^0 
\left( t, s^0, k \right)$, and $S_{V, A}^0 \left( t, s^0, k 
\right)$\footnote{The retarded and advanced propagators, e.~g.~$G_R \left(
x^0, y^0, k \right) = \theta \left( x^0 - y^0 \right) G_{\varrho} \left( x^0, 
y^0, k \right)$ and $G_A \left( x^0, y^0, k \right) = - \theta \left( y^0 - 
x^0 \right) G_{\varrho} \left( x^0, y^0, k \right)$, and self energies have to
be introduced in order to remove the upper boundaries of the memory integrals
in the \KB equations.}, as well as the corresponding self energies. In order
to be able to really perform the Fourier transformation, we have to send the
initial time to $- \infty$. At least for large $x^0$ and $y^0$ this can be
justified by taking into account that correlations between earlier and later
times are suppressed exponentially~\cite{Berges:2001fi,Lindner:2005kv}. For
early times, however, this is certainly not the case. The result of all these
transformations are quantum-kinetic equations for the statistical propagators
$G_F$ and $S_{V, F}^0$~\cite{Ivanov:1999tj,Knoll:2001jx,Blaizot:2001nr,
Berges:2002wt,Juchem:2004cs,Berges:2005vj,Berges:2005md}\footnote{The complete
and closed set of these quantum-kinetic equations comprehends 9 equations and
self energies, which, for completeness, are shown in App.~\ref{sect3.3}.}:
\begin{equation} \label{eq3.8}
  - \PB{\Omega}{G_F} = \Pi_{\varrho} G_F  - \Pi_F G_{\varrho} + \PB{\Pi_F}{\Re \left( G_R \right)}
\end{equation}
and
\begin{eqnarray}
        \lefteqn{\PB{W}{S_{V, F}^0} = \Sigma_{V, \varrho}^0 S_{V, F}^0 - \Sigma_{V, F}^0 S_{V, \varrho}^0 + \Sigma_{V, \varrho} S_{V, F}} \nonumber \\
  &   & {} - \Sigma_{V, F} S_{V, \varrho} - \PB{\Sigma_{V, F}^0}{\Re \left( S_{V, R}^0 \right)} \label{eq3.9} \\
  &   & {} + \PB{\Re \left( \Sigma_{V, R} \right)}{S_{V, F}} + \PB{\Sigma_{V, F}}{\Re \left( S_{V, R} \right)} \nonumber \;,
\end{eqnarray}
where the Poisson brackets are defined by
\[ \PB{f}{g} = - \Big[ \partial_t f \Big] \Big[ \partial_{\omega} g \Big] + \Big[ \partial_{\omega} f \Big] \Big[ \partial_t g \Big] \;. \]
The auxiliary functions
\[ \Omega \left( t, \omega, k \right) = - \omega^2 + k^2 + M^2 \left( t \right) + \Re \left( \Pi_R \left( t, \omega, k \right) \right) \]
and 
\[ W \left( t, \omega, k \right) = \omega + \Re \left( \Sigma_{V, R}^0 \left( t, \omega, k \right) \right) \]
have been introduced to simplify the notation. Employing the first-order
Taylor expansion is clearly not justifiable for early times when the
equal-time propagator is rapidly 
oscillating~\cite{Berges:2000ur,Lindner:2005kv}. Consequently, one might
expect that the
above quantum-kinetic equations and also the Boltzmann equations, which we
derive subsequently, fail to describe the early-time evolution and that errors
accumulated for early times cannot be remedied at late times. In fact, the
first-order gradient expansion is motivated by equilibrium considerations:
In equilibrium the propagator depends on the relative coordinates only. There
is no dependence on the center coordinates, and one may hope that there are
situations where the propagator depends only moderately on the center
coordinates. This is clearly the case for late times when our system is
sufficiently close to equilibrium. However, already after moderate times the
rapid oscillations mentioned above, have died out and are followed by a
monotonous drifting regime~\cite{Berges:2001fi,Lindner:2005kv}. In this
drifting regime the second derivative with respect to $t$ should be negligible
as compared to the first-order derivative, and a consistent Taylor expansion
can be justified even though the system may still be far from equilibrium.
Here, it is crucial that the Taylor expansion is performed consistently for
two reasons: First, this guarantees that the quantum-kinetic equations satisfy
exactly the same conservation laws as the full \KB equations 
do~\cite{Knoll:2001jx}. Second, it has been shown that neglecting the Poisson
brackets severely restricts the range of validity of the quantum-kinetic 
transport equations~\cite{Berges:2005ai,Berges:2005md}.

In order to derive Boltzmann equations from the quantum-kinetic equations for
the statistical propagators (\ref{eq3.8}) and (\ref{eq3.9}), first we have to
discard the Poisson brackets on the right-hand sides, thereby sacrificing the
consistency of the gradient expansion. On the left-hand sides we remove the
time dependence of the auxiliary quantities $\Omega$ and $W$. We take
\[ W \left( t, \omega, k \right) = \omega \]
and 
\[ \Omega \left( t, \omega, k \right) = - \omega^2 + k^2 + m^2 \;, \]
where $m$ is the thermal mass of the scalars. After that, we employ the \KB
ansatz
\begin{equation} \label{eq3.10} 
  G_F \left( t, \omega, k \right) = G_{\varrho} \left( \omega, k \right) \left( \frac{1}{2} + n_s \left( t, \omega, k \right) \right)
\end{equation}
and
\begin{equation} \label{eq3.11} 
  S_F \left( t, \omega, k \right) = S_{\varrho} \left( \omega, k \right) \left( \frac{1}{2} - n_f \left( t, \omega, k \right) \right)
\end{equation}
which also can be motivated by equilibrium considerations. In fact, this is a 
generalization of the fluctuation-dissipation theorem, which states that, for 
a system in thermal equilibrium, the statistical propagator is proportional to
the spectral function. The fluctuation dissipation theorem can be recovered 
from Eqs.~(\ref{eq3.10}) and (\ref{eq3.11}) by discarding the dependence on 
the center time $t$ and fixing $n_s$ and $n_f$ to be the Bose-Einstein and
Fermi-Dirac distribution function, respectively. The last approximation, which 
is necessary to arrive at the Boltzmann equations, is the so-called 
quasi-particle (or on-shell) approximation. For the scalars this means that
the spectral function takes the form
\[ G_{\varrho} \left( \omega, k \right) = \frac{\pi}{E \left( k \right)} \Big( \delta \left( \omega - E \left( k \right) \right) - \delta \left( \omega + E \left( k \right) \right) \Big) \;, \]
where the quasi-particle energy is given by
\[ E \left( k \right) = \sqrt{m^2 + k^2} \;. \]
For the lepton spectral function we assume
\begin{eqnarray*}
        S_{V, \varrho}^{0} \left( \omega, k \right)
  & = & \pi \left( \delta \left( \omega - k \right) + \delta \left( \omega + k \right) \right) \;, \\
        S_{V, \varrho} \left( \omega, k \right)
  & = & \pi \left( \delta \left( \omega - k \right) - \delta \left( \omega + k \right) \right) \;.
\end{eqnarray*}
Once more, we would like to stress that the exact time evolution of the 
spectral functions is determined by the \KB equations. It has been 
shown that the spectral function can be parameterized by a Breit-Wigner 
function with a non-vanishing width~\cite{Aarts:2001qa,Juchem:2003bi}. To 
reduce the width of this Breit-Wigner curve to zero is certainly not a 
controllable approximation and leads to very large qualitative discrepancies 
between the results produced by \KB and Boltzmann equations. In fact 
this approximation can only be justified if our system consists of stable, or 
at least very long-lived, quasi-particles, whose mass is much larger than 
their decay width. We also would like to note that a completely 
self-consistent determination of the thermal mass in the framework of the 
Boltzmann equation requires the solution of an integral equation for
$E \left( k \right)$, which would drastically increase the complexity
of our numerics. As none of our physical results depend on the exact value of
the thermal mass, for convenience, we use the equilibrium value of the thermal
scalar mass as determined by the \KB equations. Eventually, we define the
quasi-particle number densities by
\[ n_s \left( t, k \right) = n_s \left( t, E \left( k \right), k \right) \]
and
\[ n_f \left( t, k \right) = n_f \left( t, k, k \right) \;. \]
After equating the positive energy components in Eqs.~(\ref{eq3.8}) and
(\ref{eq3.9}) we arrive at the following Boltzmann equations:
\begin{eqnarray}
        \lefteqn{\partial_t n_s \left( t, k \right) = 2 \pi \eta^2 \int \frac{\ddd{p}}{\left( 2 \pi \right)^3} \int \ddd{q} \; \delta^3 \left( \bm{k} - \bm{p} - \bm{q} \right)} \; \nonumber \\
  &   & {} \times \delta \left( E \left( k \right) - p - q \right) \frac{1}{E(k)} \left( 1 - \frac{\bm{p} \bm{q}}{pq} \right) \nonumber \\
  &   & {} \times \bigg[ \Big( n_s \left( t, k \right) + 1 \Big) n_f \left( t, p \right) n_f \left( t, q \right) \nonumber \\
  &   & {} - n_s \left( t, k \right) \Big( n_f \left( t, p \right) - 1 \Big) \Big( n_f \left( t, q \right) - 1 \Big) \bigg] \label{eq3.24}
\end{eqnarray}
and
\begin{eqnarray}
        \lefteqn{\partial_t n_f \left( t, k \right) = 2 \pi \eta^2 \int \frac{\ddd{p}}{\left( 2 \pi \right)^3} \int \ddd{q} \; \delta^3 \left( \bm{k} + \bm{p} - \bm{q} \right)} \; \nonumber \\
  &   & {} \times \delta \left( k + p - E \left( q \right) \right) \frac{1}{E(q)} \left( 1 - \frac{\bm{k} \bm{p}}{kp} \right) \nonumber \\
  &   & {} \times \bigg[ \Big( n_f \left( t, k \right) - 1 \Big) \Big( n_f \left( t, p \right) - 1 \Big) n_s \left( t, q \right) \nonumber \\
  &   & {} - n_f \left( t, k \right) n_f \left( t, p \right) \Big( n_s \left( t, q \right) + 1 \Big) \bigg] \;. \label{eq3.25}
\end{eqnarray}
Exploiting isotropy the above 6-dimensional Boltzmann collision integrals can
be reduced to 1-dimensional integrals:
\begin{eqnarray}
        \lefteqn{\partial_t n_s \left( t, k \right) = \frac{2 \eta^2}{\pi^2 E \left( k \right)} \intl_0^{\infty} dp \; \theta \left( q_0 \right) J_s \left( k, p, q_0 \right)} \nonumber \\
  &   & {} \times \bigg[ \Big( n_s \left( t, k \right) + 1 \Big) n_f \left( t, p \right) n_f \left( t, q_0 \right) \label{eq3.29} \\
  &   & {} - n_s \left( t, k \right) \Big( n_f \left( t, p \right) - 1 \Big) \Big( n_f \left( t, q_0 \right) - 1 \Big) \bigg] \;, \nonumber
\end{eqnarray}
\begin{eqnarray}
        \lefteqn{\partial_t n_f \left( t, k \right) = \frac{2 \eta^2}{\pi^2} \intl_0^{\infty} dq \; \theta \left( p_0 \right) \; \frac{q}{E \left( q \right)} J_f \left( k, p_0, q \right)} \nonumber \\
  &   & {} \times \bigg[ \Big( n_f \left( t, k \right) - 1 \Big) \Big( n_f \left( t, p_0 \right) - 1 \Big) n_s \left( t, q \right) \nonumber \\
  &   & {} - n_f \left( t, k \right) n_f \left( t, p_0 \right) \Big( n_s \left( t, q \right) + 1 \Big) \bigg] \;. \label{eq3.30}
\end{eqnarray}
The details of these calculations and the definitions of all of the auxiliary
quantities are given in App.~\ref{sect3.4}. This simplification of the
collision integrals is crucial in order to implement efficient computer
programs for the numerical solution of the Boltzmann equations.

In this section we showed that the derivation of Boltzmann equations from \KB
equations requires a number of non-trivial approximations and assumptions. One
has to employ a first-order gradient expansion, a Wigner transformation and a
quasi-particle approximation. In this sense, one can consider the \KB
equations as quantum Boltzmann equations re-summing the gradient expansion up
to infinite order and including memory and off-shell effects.

\section{Comparing Boltzmann vs.~\KB \label{sect3.1}}

\subsection{Initial Conditions and Numerical Settings}

\begin{figure*}
  \centering
  \hspace*{\fill}
  \includegraphics[height=55mm]{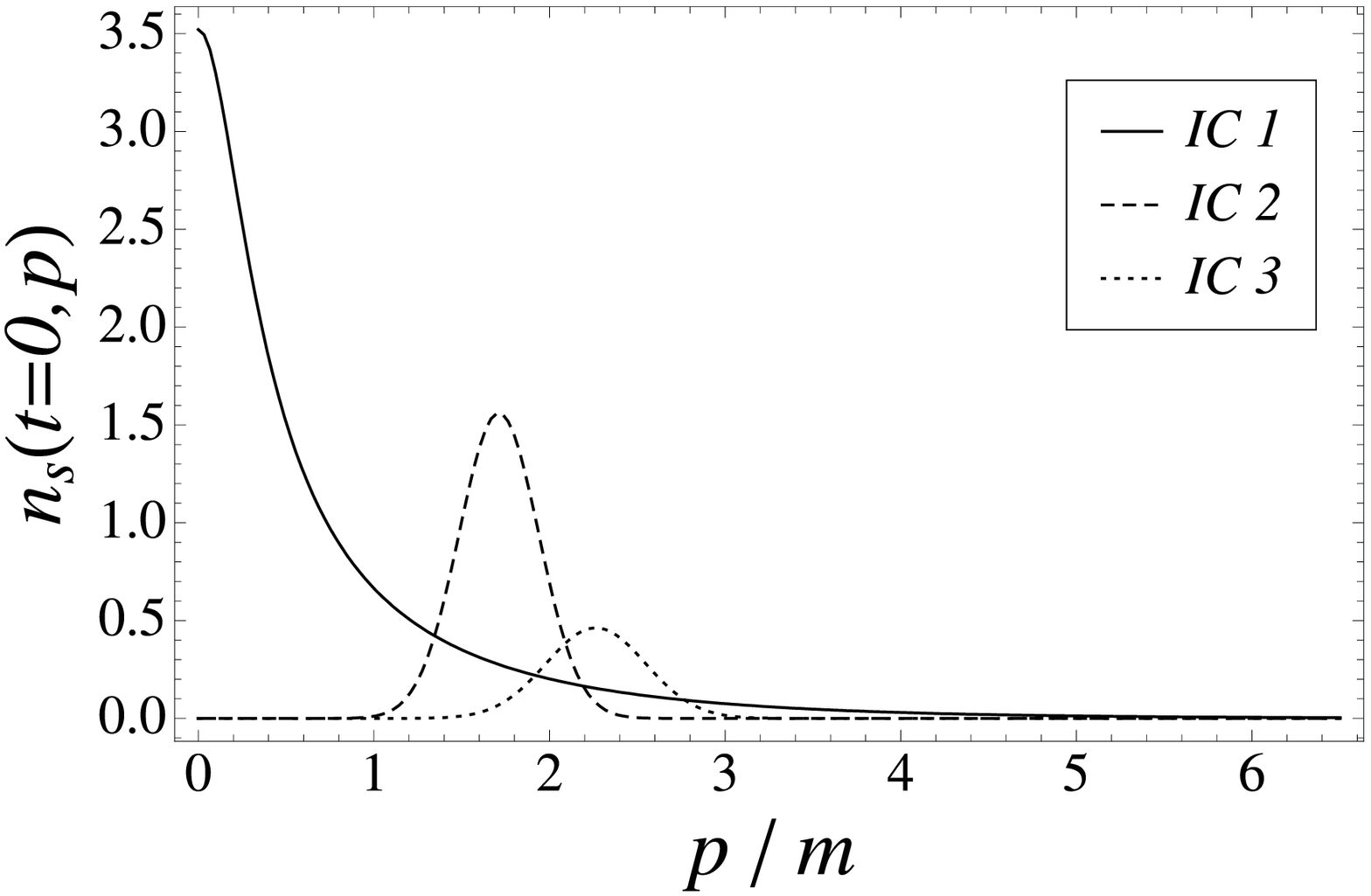}
  \hfill \hfill
  \includegraphics[height=55mm]{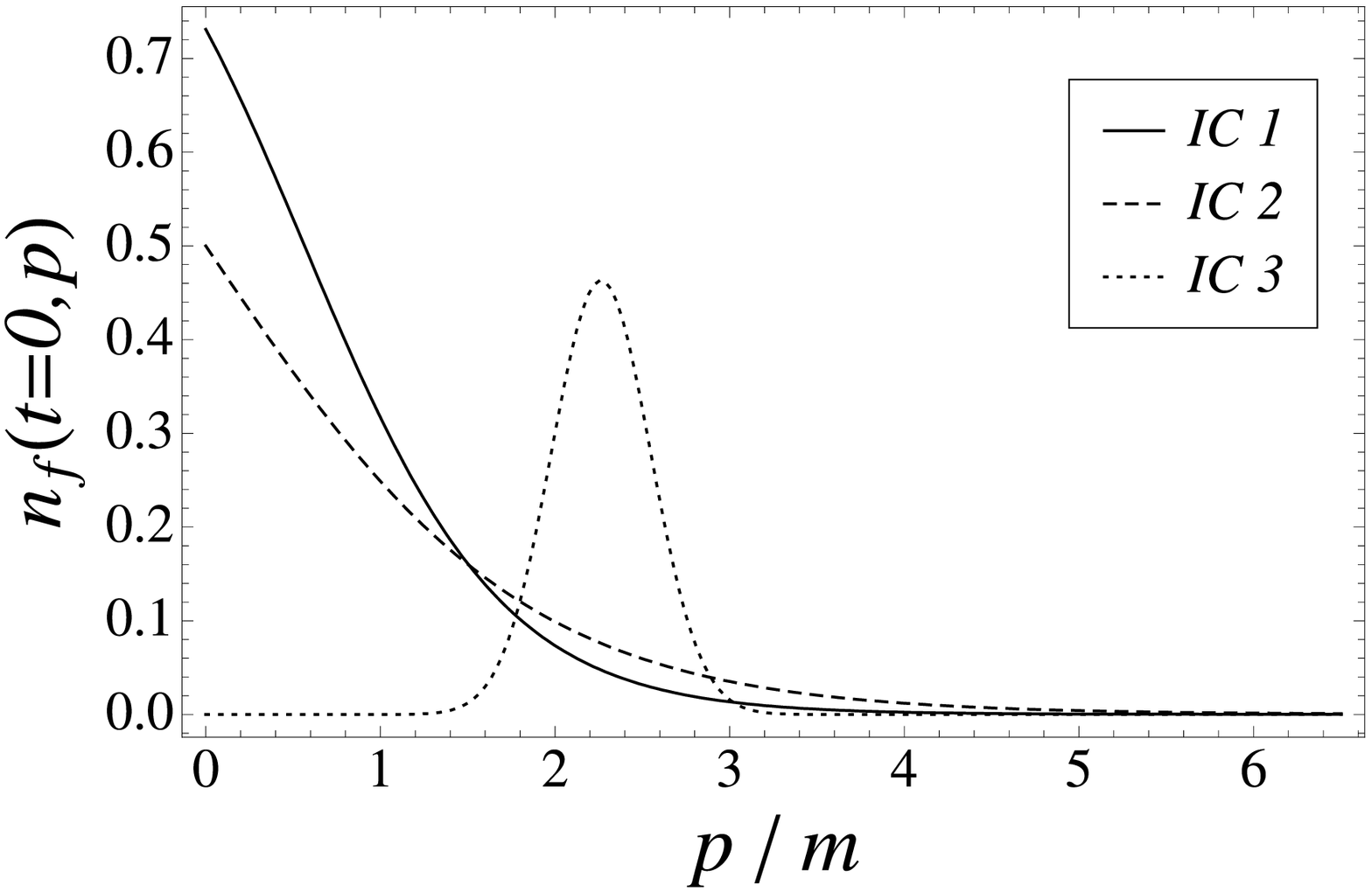}
  \hspace*{\fill}
  \caption{\label{fig3.5}Initial particle number distributions for scalars
  (left) and fermions (right). These distributions define the initial 
  conditions (IC) for which we numerically solved the Boltzmann and
  \KB equations. All initial conditions correspond to the same
  (conserved) average energy density. Above that, for the initial conditions
  IC1 and IC2 also the sum of the initial average number densities agree.}
\end{figure*}
In order to solve the \KB equations numerically, we follow exactly
the lines of Refs.~\cite{Berges:2002wr,Lindner:2005kv,montvayMunster1994a} on
a lattice with $2000^2 \times 32^3$ lattice sites. The values for the coupling
constants are $\eta = 1$ and $\lambda = 0.25$. The initial conditions for
the statistical propagators are determined by scalar and fermionic particle
number distributions $n_s \left( t=0, p \right)$ and $n_f \left( t=0, p
\right)$ according to
\begin{eqnarray*}
        G_F \left( x^0, y^0, p \right)_{x^0=y^0=0} 
  & = & \frac{n_s \left( t=0, p \right) + \frac{1}{2}}{\omega_0 \left( p \right)} \;, \\
        \left[ \partial_{x^0} G_F \left( x^0, y^0, p \right) \right]_{x^0=y^0=0}
  & = & 0 \;,
\end{eqnarray*}
\begin{eqnarray*}
  \lefteqn{\left[ \partial_{x^0} \partial_{y^0} G_F \left( x^0, y^0, p \right) \right]_{x^0=y^0=0}} \; \\
  & = & \omega_0 \left( p \right) \left( n_s \left( t=0, p \right) + \frac{1}{2} \right) \;,
\end{eqnarray*}
\begin{eqnarray*}
        S_{V, F}^0 \left( x^0, y^0, p \right)_{x^0=y^0=0} 
  & = & 0 \;, \\
        S_{V, F} \left( x^0, y^0, p \right)_{x^0=y^0=0} 
  & = & \frac{1}{2} - n_f \left( t=0, p \right) \;,
\end{eqnarray*}
where $\omega_0 \left( p \right)$ is the initial scalar kinetic energy
distribution. On the other hand, the initial conditions for the spectral
functions are determined by equal-time \mbox{(anti-)} commutation relations.
We solve the Boltzmann and \KB equations for three different sets of initial
particle number distributions, which are shown in Figure~\ref{fig3.5}. All
initial conditions correspond to the same (conserved) average energy density.
Above that, for the initial conditions IC1 and IC2 also the sum of the initial
scalar and fermionic average particle number densities agree. The numerical
solution of the Boltzmann equations proceeds along the lines of
Ref.~\cite{Lindner:2005kv}. The scale in our plots is set by the scalar
thermal mass $m = \omega_{eq} \left( p=0 \right)$, where $\omega_{eq} \left( 
p \right)$ is the effective kinetic energy distribution (\ref{eq3.47}) for
sufficiently late time $t$.

\subsection{Universality}

\begin{figure*}
  \centering
  \hspace*{\fill}
  \includegraphics[height=60mm]{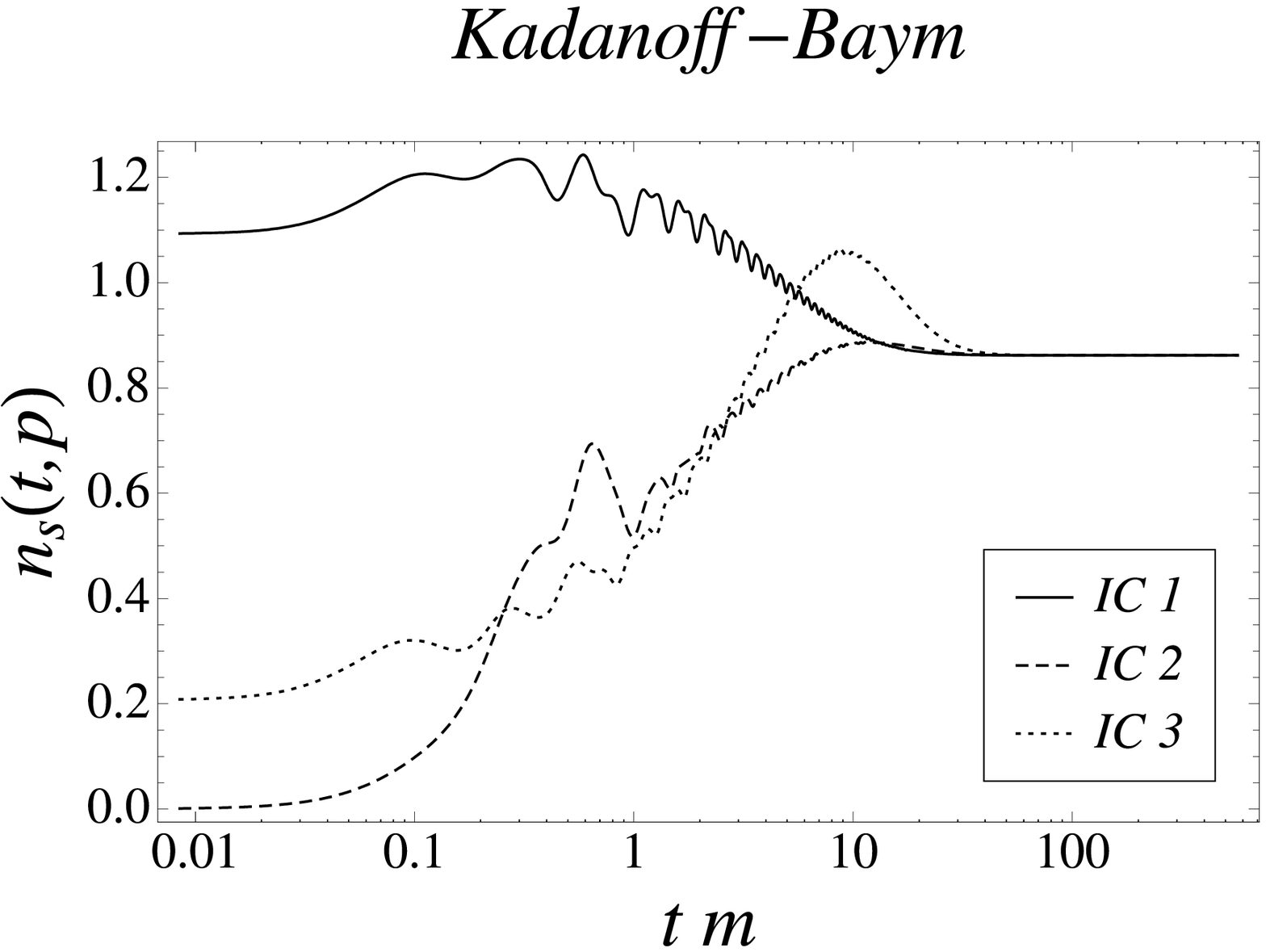}
  \hfill \hfill
  \includegraphics[height=60mm]{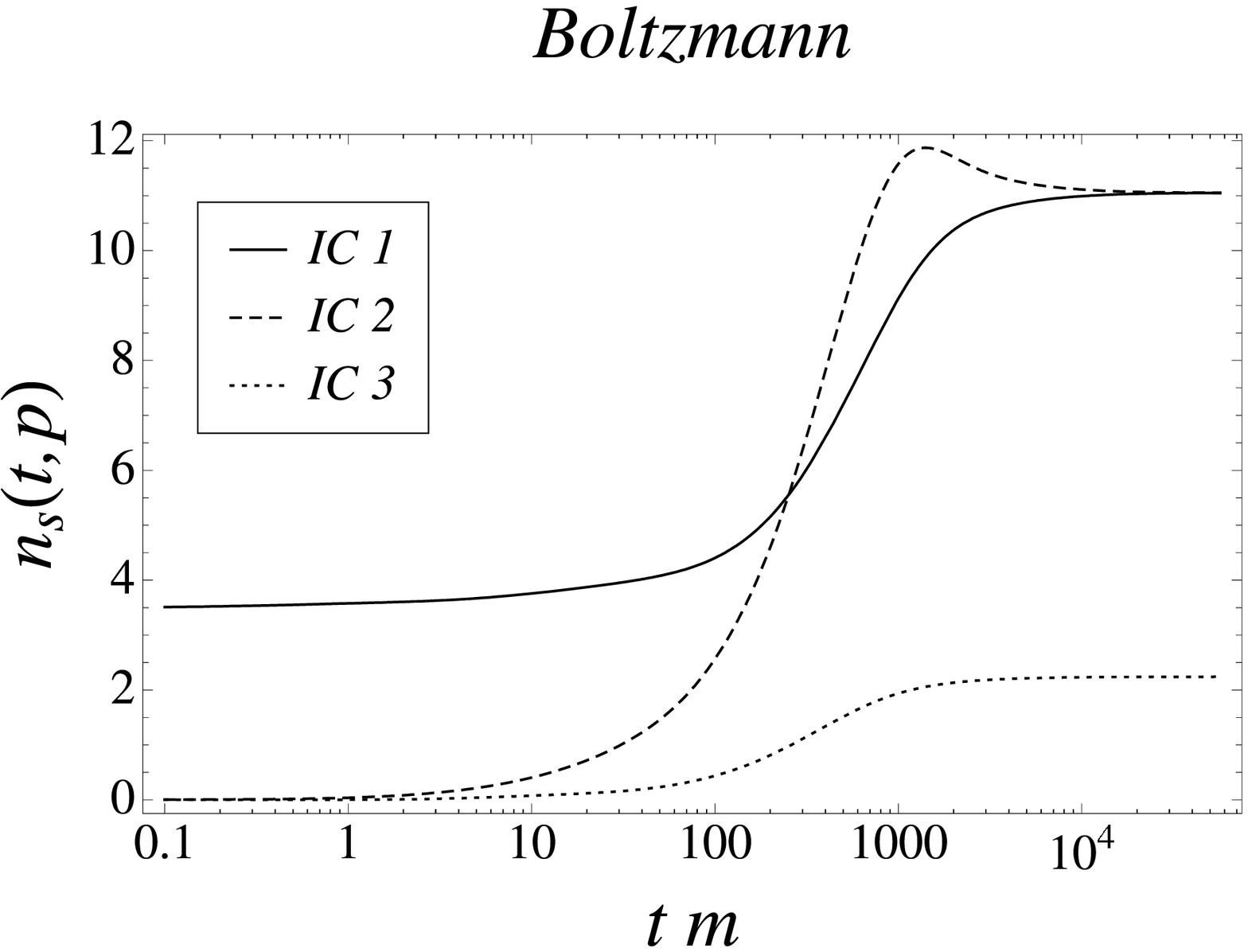}
  \hspace*{\fill} \\
  \hspace*{\fill}
  \includegraphics[height=55mm]{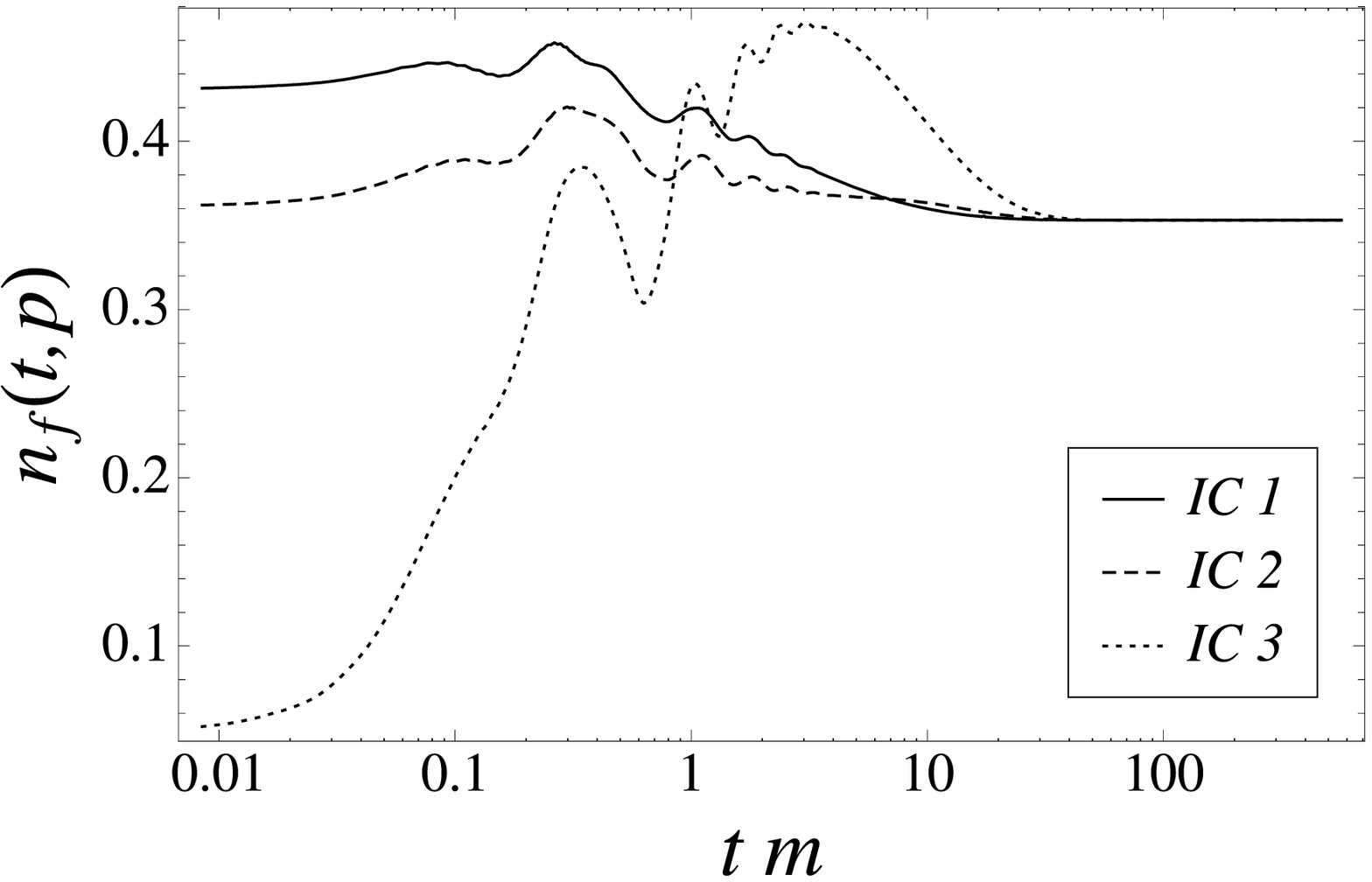}
  \hfill \hfill
  \includegraphics[height=55mm]{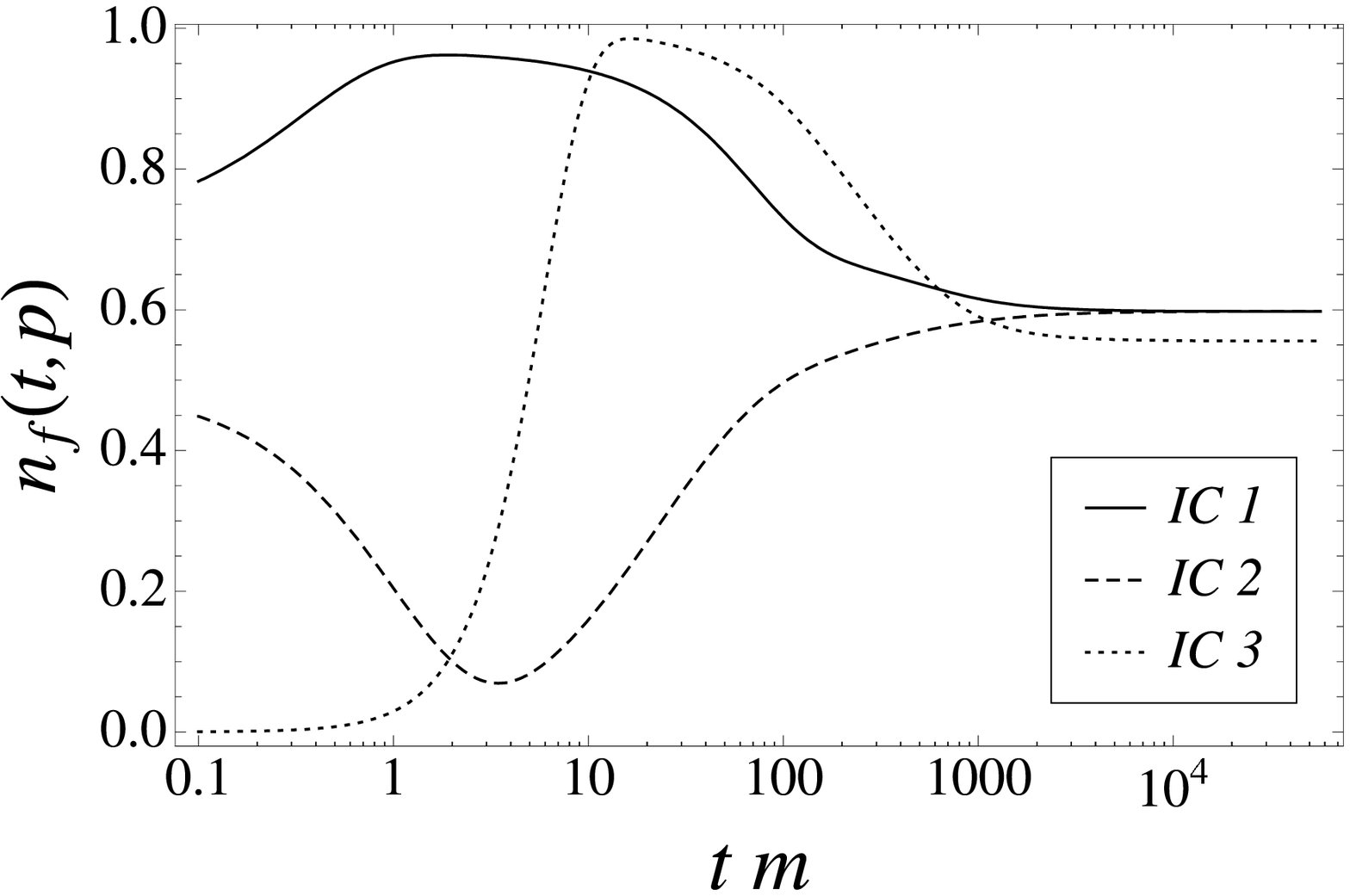}
  \hspace*{\fill}
  \caption{\label{fig3.6}Time evolution of the particle number distributions
  for a fixed momentum mode $p$. We see that \KB equations respect full
  universality, whereas Boltzmann equations maintain only a restricted
  universality.}
\end{figure*}

\begin{figure*}
  \centering
  \hspace*{\fill}
  \includegraphics[height=55mm]{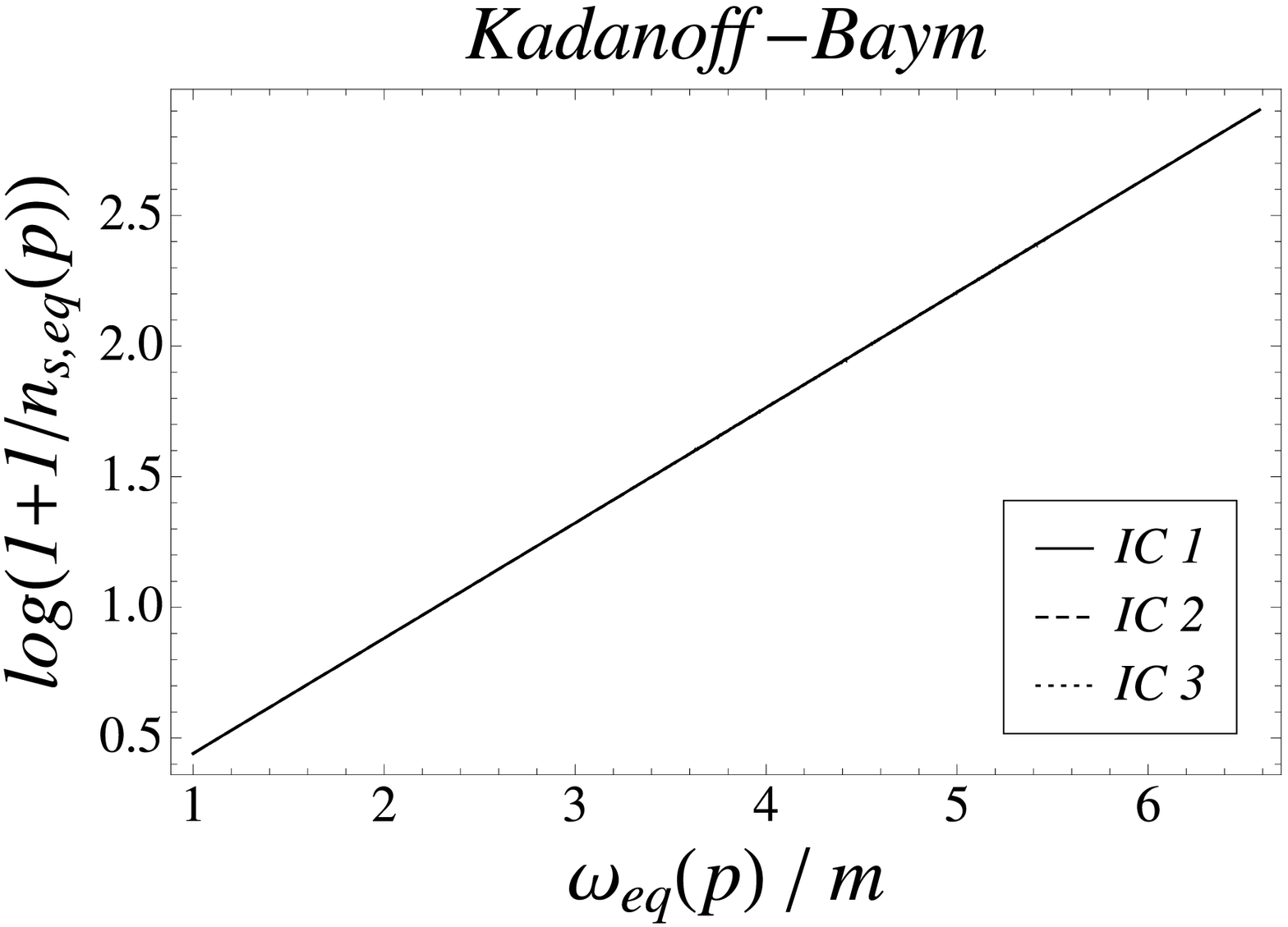}
  \hfill \hfill
  \includegraphics[height=55mm]{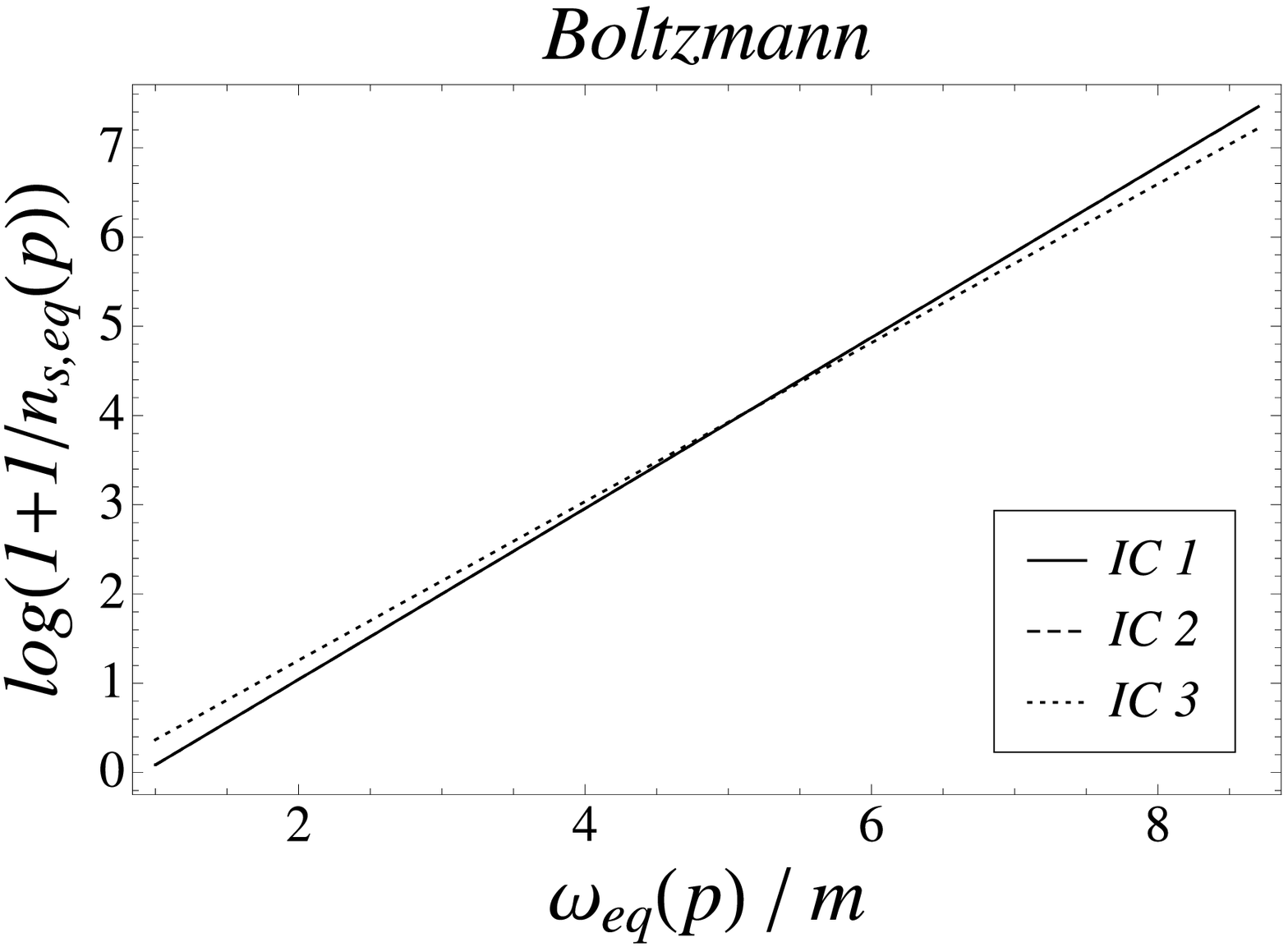}
  \hspace*{\fill} \\
  \hspace*{\fill}
  \includegraphics[height=50mm]{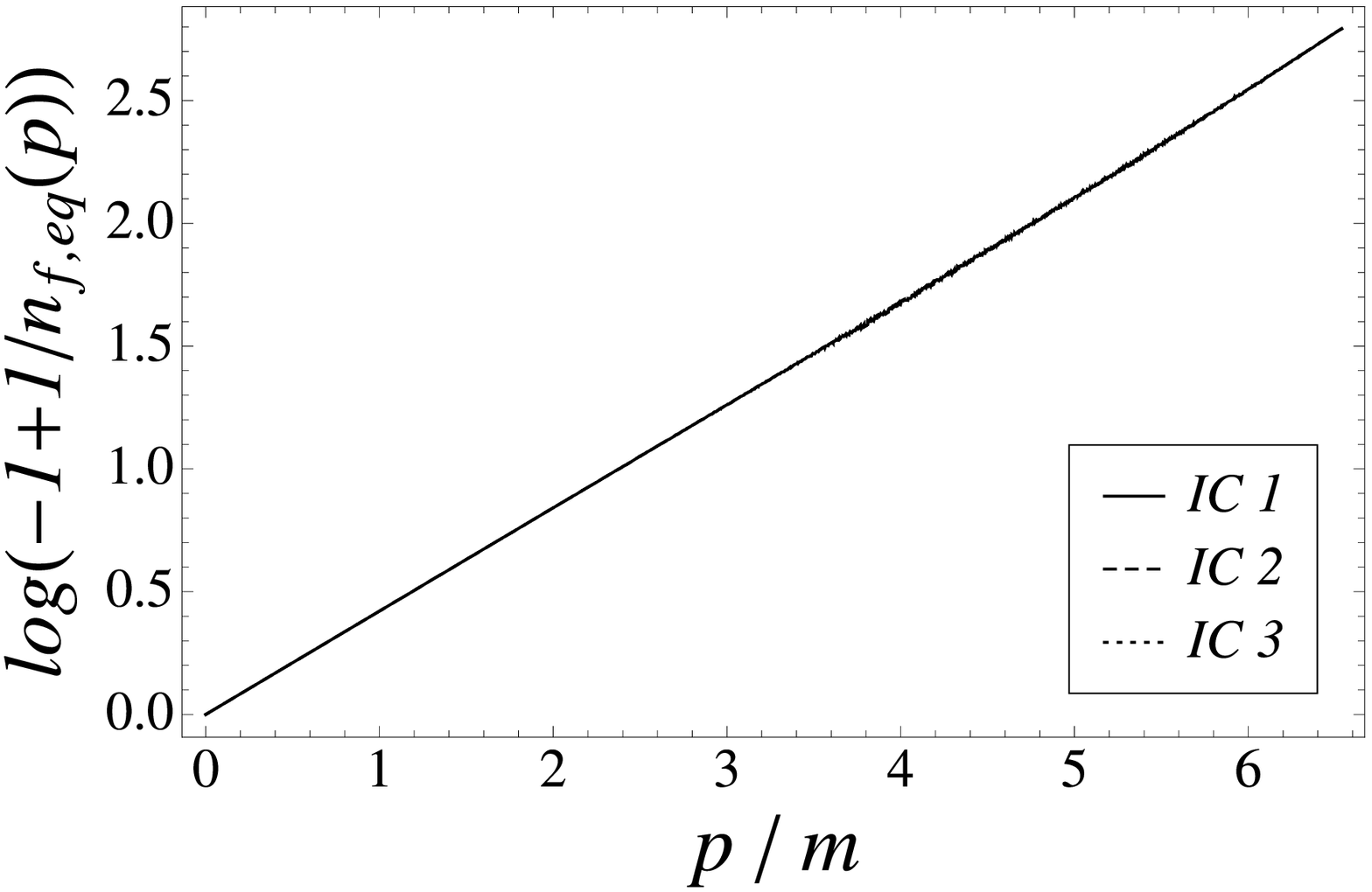}
  \hfill \hfill
  \includegraphics[height=50mm]{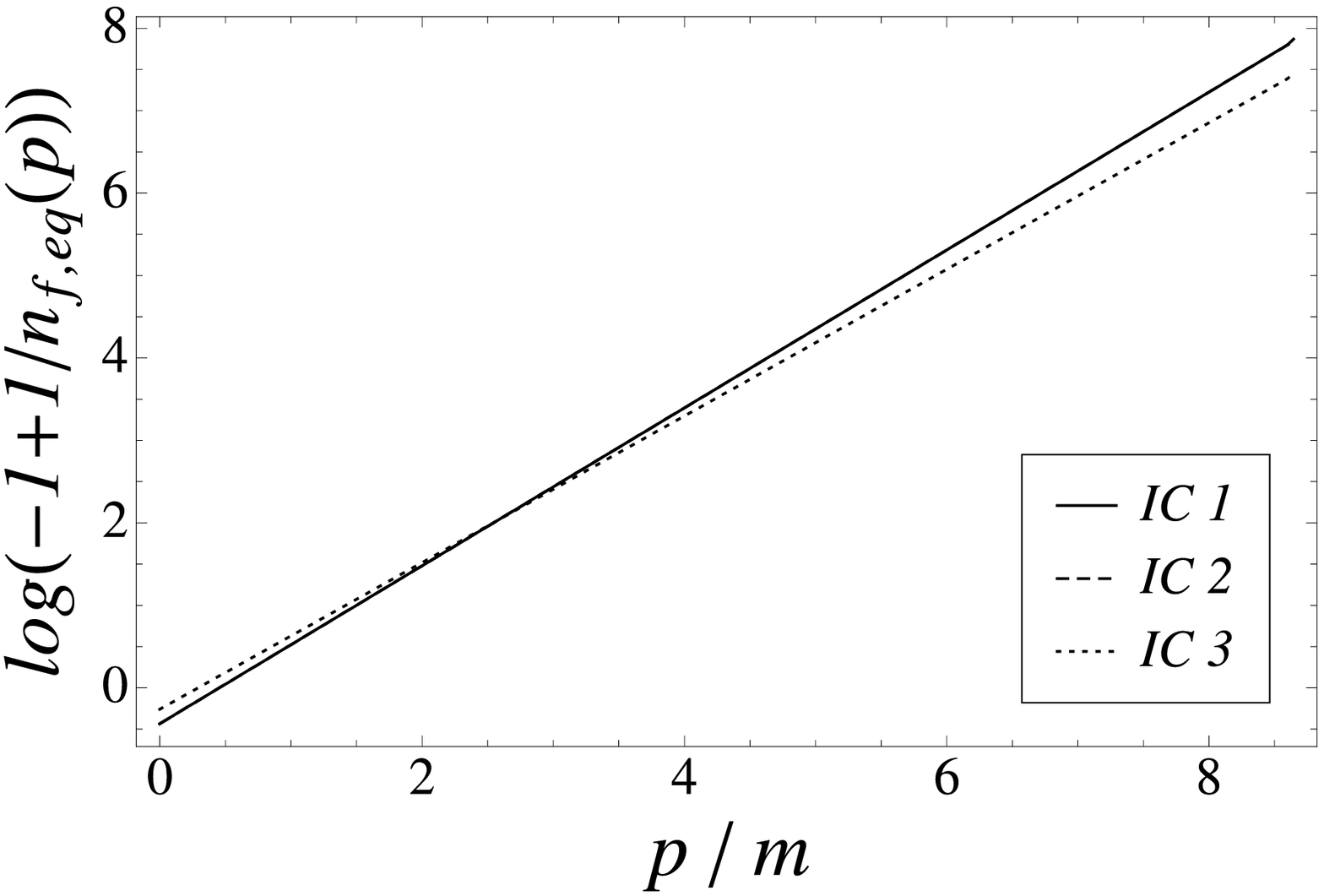}
  \hspace*{\fill}
  \caption{\label{fig3.9}Equilibrium particle number distributions. The
  straight lines in these plots prove that in equilibrium the particle number
  distributions indeed take the form of Bose-Einstein or Fermi-Dirac
  distribution functions, where the temperature is given by the inverse slope
  and the chemical potential is proportional to the y-axis intercept. The \KB 
  equations lead to a universal temperature $T = 2.3 m$ and universally
  vanishing chemical potentials. The temperatures and chemical potentials
  predicted by the Boltzmann equations are given in Table~\ref{tab3.1}.}
\end{figure*}

Figs.~\ref{fig3.6} and \ref{fig3.9} exhibit that the \KB equations
respect full universality: Figure~\ref{fig3.6} shows the time evolution of the
particle number distributions for a fixed momentum mode. The particle
number distributions start from different initial values and go through very
different early-time evolutions. Nevertheless, in the case of the 
\KB equations they all approach the same universal late-time value.
Figure~\ref{fig3.9} shows the particle number distributions for times when
equilibrium has effectively been reached. In the case of \KB equations, we
observe that the equilibrium number distributions agree exactly independent of
the initial conditions, which proves that we could have shown the plots of
Figure~\ref{fig3.6} for any momentum mode. In particular, the straight lines in
Figure~\ref{fig3.9} prove that the equilibrium number distributions take the
form of Bose-Einstein or Fermi-Dirac distribution functions with a universal
temperature $T = 2.3 m$ and universally vanishing chemical potentials.

In contrast to this, Boltzmann equations maintain only a restricted
universality. Figs.~\ref{fig3.6} and \ref{fig3.9} reveal that only the
initial conditions IC1 and IC2 lead to the same late-time behavior, which
deviates significantly from the one approached by the third initial condition
IC3. Again, the straightness of the lines in Figure~\ref{fig3.9} proves that
the equilibrium number distributions take the Form of Bose-Einstein or
Fermi-Dirac distribution functions. However, the different slopes of the lines
indicate that the temperature is not the same for all initial conditions, and
the non-vanishing y-axis intercepts indicate that Boltzmann equations may
predict different non-vanishing chemical potentials. Fitted values for these
quantities are given in Table~\ref{tab3.1}.

\begin{figure*}
  \centering
  \hspace*{\fill}
  \includegraphics[height=60mm]{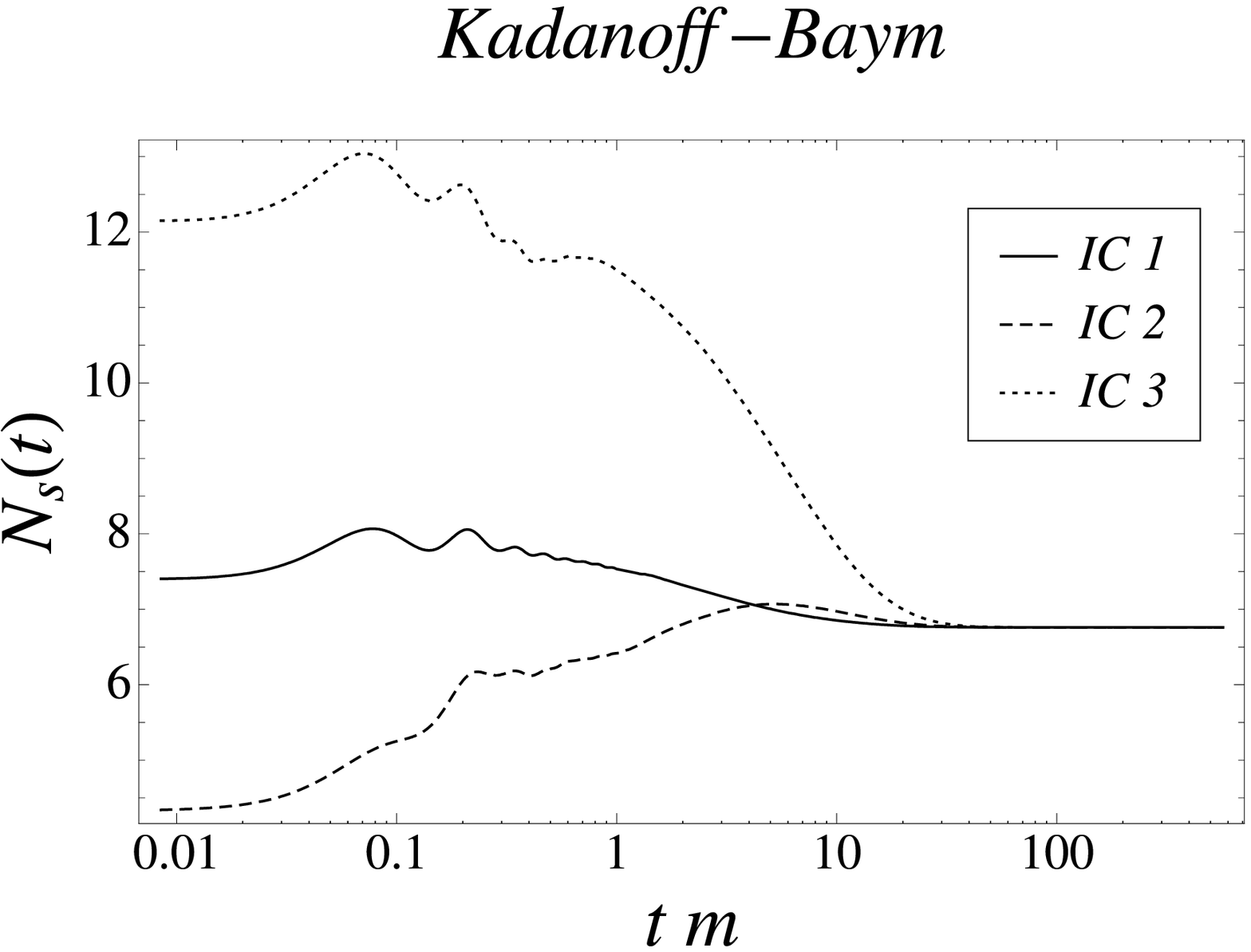}
  \hfill \hfill
  \includegraphics[height=60mm]{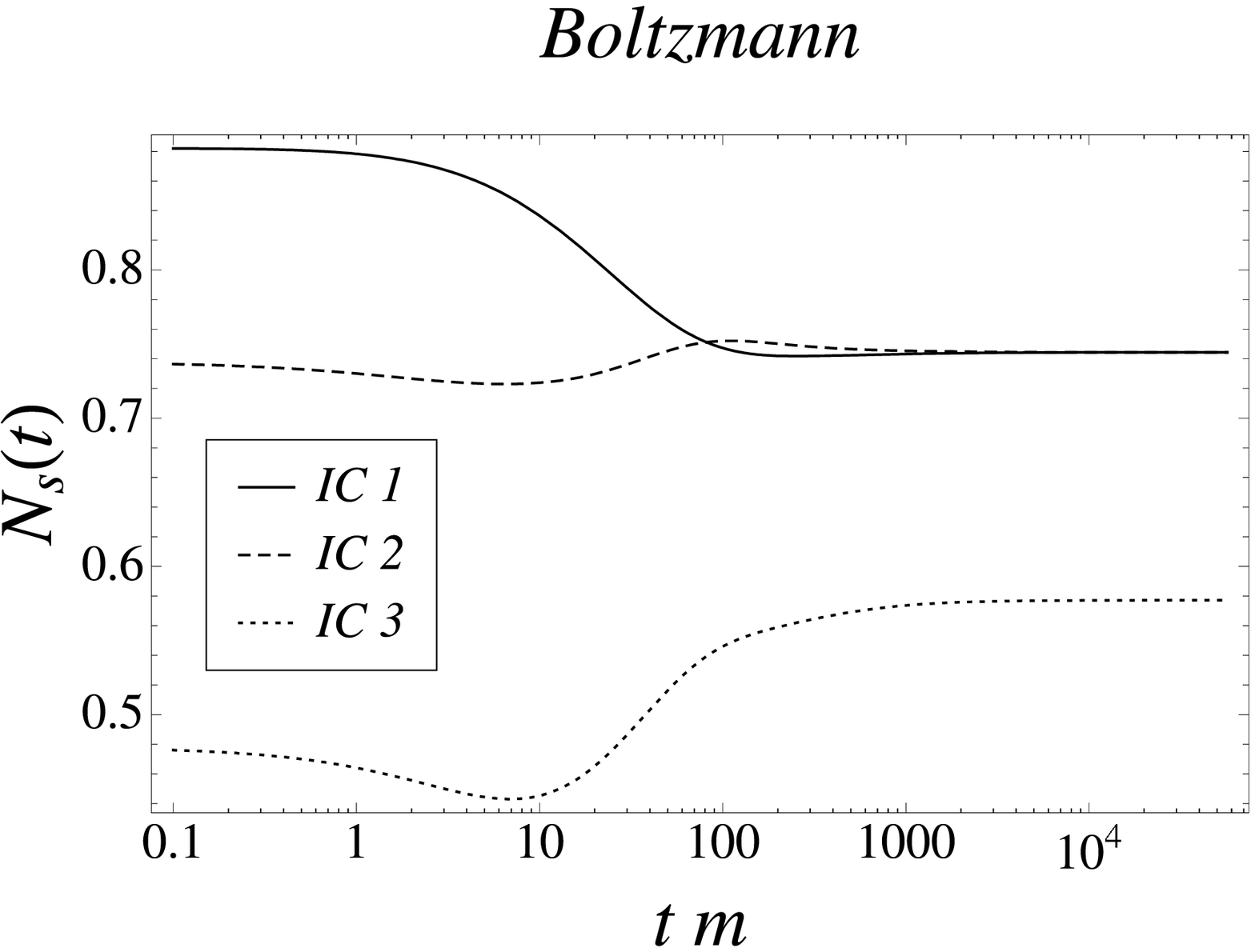}
  \hspace*{\fill} \\
  \hspace*{\fill}
  \includegraphics[height=55mm]{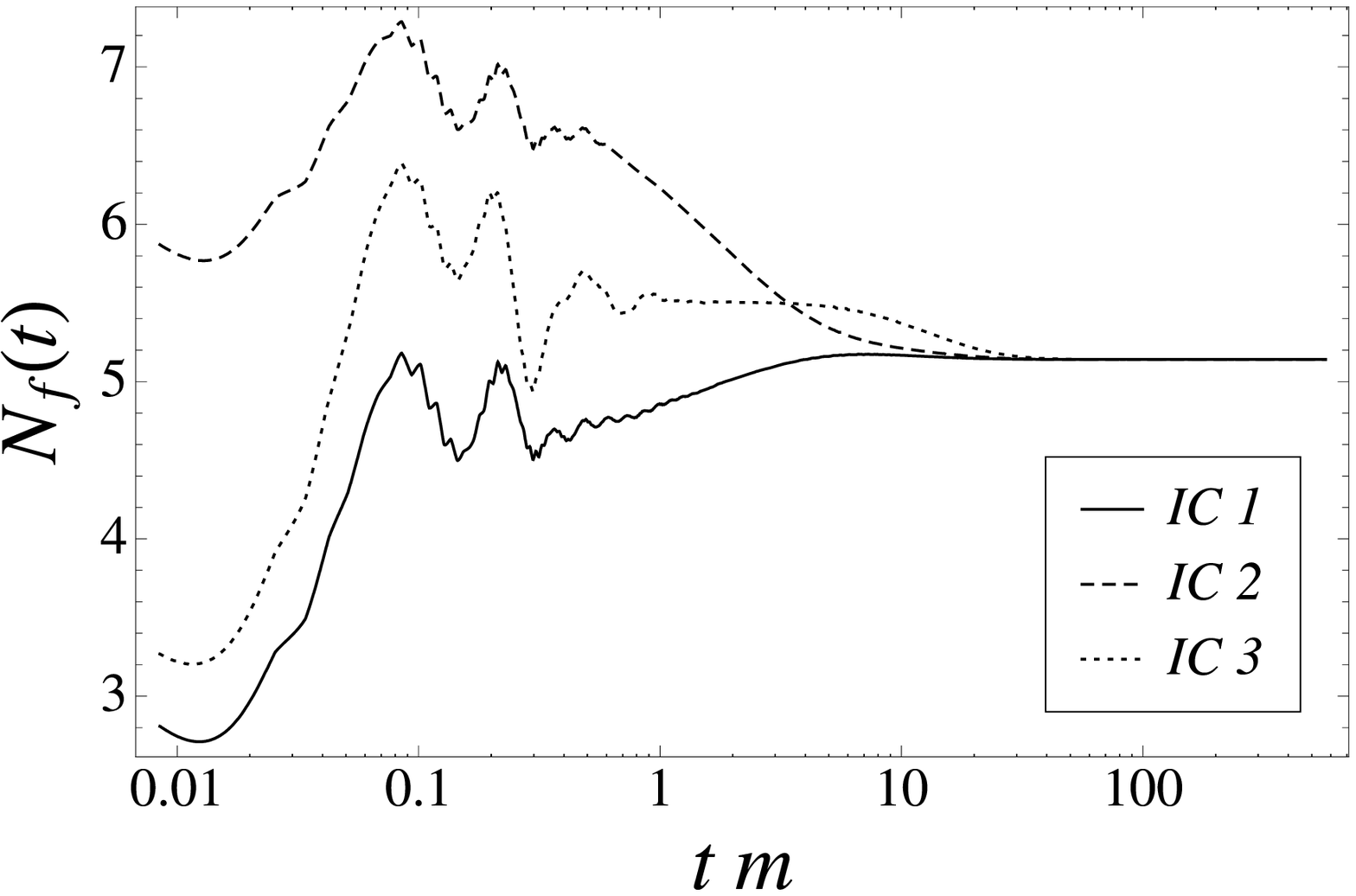}
  \hfill \hfill
  \includegraphics[height=55mm]{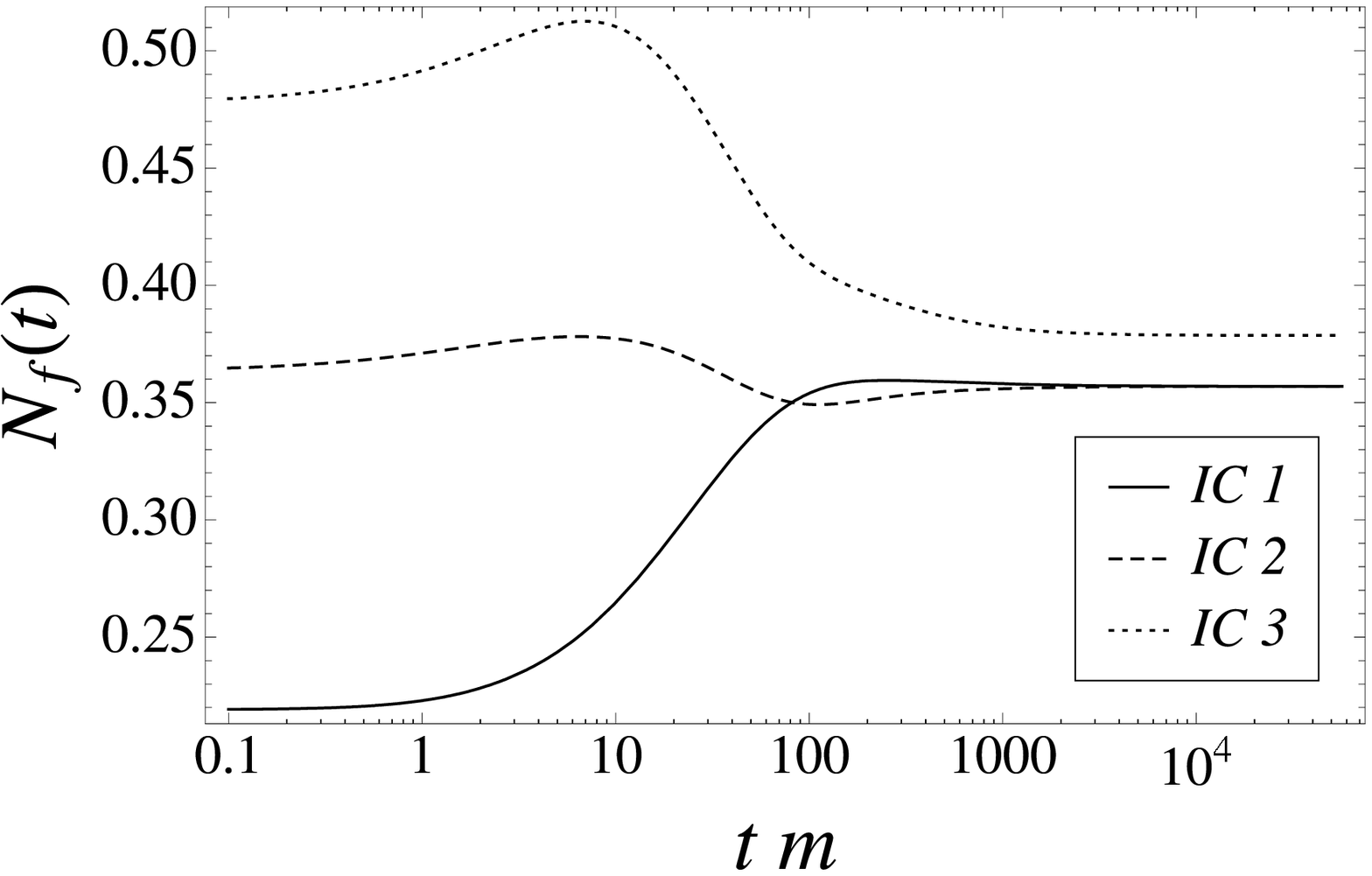}
  \hspace*{\fill} \\
  \hspace*{\fill}
  \includegraphics[height=55mm]{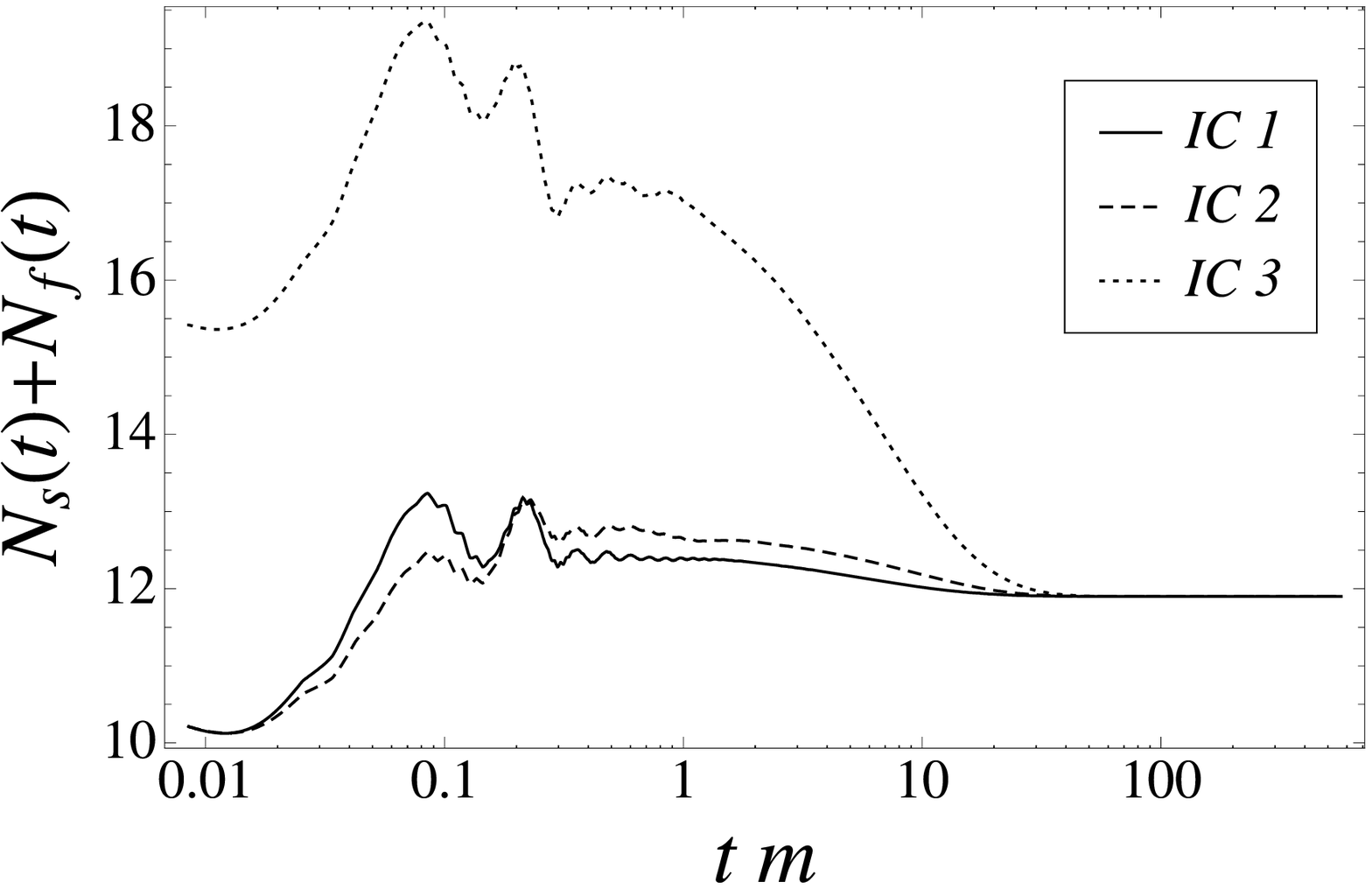}
  \hfill \hfill
  \includegraphics[height=55mm]{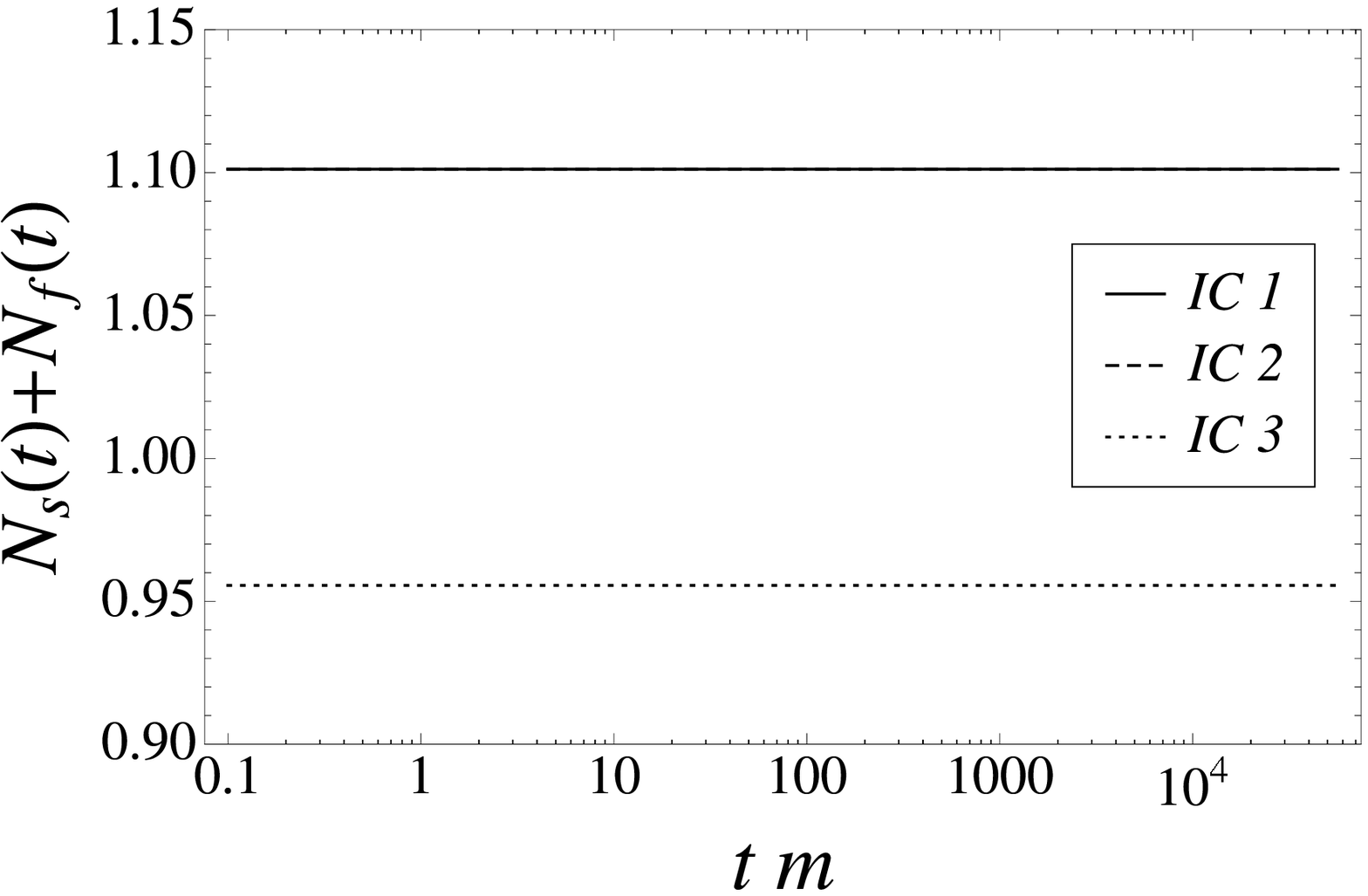}
  \hspace*{\fill}
  \caption{\label{fig3.7}Time evolution of the average particle number
  densities and their sum. \KB equations allow for a change of the latter
  quantity, whereas it is artificially conserved by Boltzmann equations.
  The quantitative disagreement of the average particle number densities can
  be attributed to different initial conditions and discretization schemes
  for the Boltzmann and the \KB equations, and is of no relevance for the
  purpose of the present work.}
\end{figure*}

The reason for the observed restriction of universality can be extracted from
Figure~\ref{fig3.7} where we plotted the time evolution of the average particle
number densities per degree of freedom
\begin{equation} \label{eq3.32}
  N_s \left( t \right) = \int \frac{\ddd{p}}{\left( 2 \pi \right)^3} \; n_s \left( t, p \right)
\end{equation}
and 
\begin{equation} \label{eq3.33}
  N_f \left( t \right) = \int \frac{\ddd{p}}{\left( 2 \pi \right)^3} \; n_f \left( t, p \right)
\end{equation}
and their sum. Provided $\Phi$-derivable approximations are employed,
\KB equations conserve the average energy density as well as global
charges~\cite{baym1961a,Baym:1962sx,Ivanov:1998nv}. However, as we consider
systems with vanishing net charge density neither of the above average
particle number densities has to be conserved, nor their sum. Indeed,
\KB equations include off-shell particle creation and
annihilation~\cite{Lindner:2005kv,Aarts:2001qa}, so that all of the quantities
plotted in Figure~\ref{fig3.7} may change as time goes on and approach a
universal equilibrium value.

In contrast to this, due to the quasi-particle (or on-shell) approximation the
Boltzmann equations (\ref{eq3.29}) and (\ref{eq3.30}) only include decay and
recombination processes of the form
\begin{equation} \label{eq3.31}
  \mbox{1 scalar} \longleftrightarrow \mbox{2 fermions} \;.
\end{equation}
More precisely, one of four scalars may decay into or be recombined from one
of two fermion pairs. As a consequence the sum of the average particle number
densities (\ref{eq3.32}) and (\ref{eq3.33}) is strictly conserved, as can be
seen in Figure~\ref{fig3.7}. Of course, this artificial constant of motion
severely restricts the evolution of the particle number distributions. As a
result, the Boltzmann equations maintain only a restricted universality and,
as will be discussed in the next subsection, fail to describe the process of
quantum-chemical equilibration.

\subsection{Chemical Equilibration}

\begin{table}
  \[ \begin{array}{c|c|c|r|r|c}
         &  T_s / m & T_f / m & \mu_s / m & \mu_f / m & \mu_s / \mu_f \\
    \hline
    IC 1 & 1.044 & 1.044 & 0.910 & 0.455 & 1.999 \\
    IC 2 & 1.044 & 1.044 & 0.910 & 0.455 & 2.001 \\
    IC 3 & 1.125 & 1.124 & 0.586 & 0.293 & 1.997
  \end{array} \]
  \caption{\label{tab3.1}Temperatures and chemical potentials as predicted
  by the Boltzmann equations. The values in this table have been obtained
  by fitting the equilibrium particle number distributions shown in
  Figure~\ref{fig3.9} against Bose-Einstein and Fermi-Dirac distribution
  functions, respectively.}
\end{table}
In a system allowing for creation and annihilation of particles, the chemical
potential of particles, whose total number is not restricted by any conserved
quantity, must vanish in thermodynamic equilibrium. Accordingly, as we
consider systems with vanishing net charge density the chemical potentials for
scalars and fermions should vanish once equilibrium has been reached. Indeed,
\KB equations lead to universally vanishing chemical potentials. In contrast
to this, as one can see in Table~\ref{tab3.1}, the Boltzmann equations in
general will give non-vanishing chemical potentials. For a system which
includes only interactions of the form (\ref{eq3.31}), in equilibrium the
chemical potentials are expected to satisfy the relation
\[ \mu_s = 2 \mu_f \;. \]
As one can see in the right-most column of Table~\ref{tab3.1}, this relation
is indeed fulfilled up to numerical errors $< 0.3 \%$. Thus, the Boltzmann
equations (\ref{eq3.29}) and (\ref{eq3.30}) lead to a classical chemical
equilibrium. As mentioned above, however, quantum-chemical equilibrium
requires that the chemical potentials vanish for systems with vanishing net
charge density. In this sense, the non-vanishing chemical potentials in
Table~\ref{tab3.1} indicate that the description of quantum-chemical
equilibration is out of reach of the Boltzmann equations (\ref{eq3.29}) and
(\ref{eq3.30}).

\subsection{Separation of Time Scales}

As has been discussed in Ref.~\cite{Lindner:2005kv} in the framework of a
purely scalar theory and in Ref.~\cite{Berges:2004ce} in the framework of the
linear sigma model underlying our studies in this work, \KB equations
strongly separate the time scales between the kinetic and the complete
thermodynamic (including chemical) equilibration. This phenomenon has been
called prethermalization~\cite{Berges:2004ce}, and implies that certain
quantities approach their equilibrium values on time scales which are
dramatically shorter than the thermodynamic equilibration time.

As we have seen in this work and in Ref.~\cite{Lindner:2005kv}, standard
Boltzmann equations cannot describe the phenomenon of quantum-chemical
equilibration, and thus they also cannot describe the approach to the
quantum-thermodynamic equilibrium. Consequently, standard Boltzmann equations
cannot separate the time scales between the kinetic and the full thermodynamic
equilibration and hence a description of prethermalization is out of reach of
standard Boltzmann equations.

\section{Conclusions}

In this article we addressed the question how reliable Boltzmann equations are
as approximations to Kadanoff-Baym equations in the framework of a chirally
invariant Yukawa-type quantum field theory coupling scalars with fermions.
Starting from the 2PI effective action, we reviewed the derivation of the \KB
equations and the approximations which are necessary to eventually arrive at
standard Boltzmann equations. We solved the Boltzmann and \KB equations
numerically for highly symmetric systems in 3+1 space-time dimensions without
any further approximations and compared their solutions for various
non-equilibrium initial conditions.

We demonstrated that the \KB equations respect universality: For systems with
equal average energy density the late-time behavior coincides independent of
the details of the initial conditions. In particular, independent of the
initial conditions the particle number distributions, temperatures, chemical
potentials and thermal masses predicted for times, when equilibrium has 
effectively been reached, coincide. Above that, \KB equations incorporate the
process of quantum-chemical equilibration: For systems with vanishing net
charge density the chemical potentials vanish once equilibrium has effectively
been reached. Last but not least, \KB equations feature the phenomenon of
prethermalization and separate the time scales between kinetic and full
thermodynamic (including quantum-chemical) equilibration.

The quasi-particle approximation introduces spurious constants of motion for
standard Boltzmann equations (cf.~Figure~\ref{fig3.7}), which severely
restricts the evolution of the particle number distributions. As a result,
Boltzmann equations cannot lead to a universal quantum-thermal equilibrium and 
maintain only a restricted universality: Only initial conditions for which the
average energy density, all global charges and all spurious constants of
motion agree from the very beginning, lead to the same equilibrium results. As
shown in Table~\ref{tab3.1}, Boltzmann equations cannot describe the
phenomenon of quantum-chemical equilibration and, in general, will lead to
non-vanishing chemical potentials even for systems with vanishing net charge
density. Due to the lack of quantum-chemical equilibration, the separation of
time scales observed for the \KB equations is absent in the case of Boltzmann
equations, which renders the description of prethermalization impossible.

Some of the approximations, which are required to derive Boltzmann equations
from \KB equations, are clearly motivated by equilibrium considerations.
Taking the observed restriction of universality into account, we conclude that
in the context of relativistic quantum fields one can safely apply standard
Boltzmann equations only to systems which are sufficiently close to
equilibrium, so that the spurious constants of motion emerging in Boltzmann
equations already take their equilibrium values. However, for systems far from
equilibrium standard Boltzmann equations work reliably neither for early times
(no prethermalization) nor for late times (only restricted late-time
universality, no quantum-chemical equilibration). Accordingly, for systems
in the intermediate regime the results given by standard Boltzmann equations
should be treated with care. For realistic scenarios, like leptogenesis or the
quark-gluon plasma, non-negligible corrections to Boltzmann equations are
expected, which should be evaluated.

Solving \KB equations numerically is significantly more difficult than solving
the corresponding standard Boltzmann equations. However, the considerable
discrepancies found for numerical solutions of \KB and Boltzmann equations
revealed equally significant limitations for standard Boltzmann equations.
Accordingly, the importance of numerical solutions of \KB equations cannot be
over-estimated and it is certainly worth to face the arising difficulties.

In the present work we considered standard Boltzmann equations at lowest order
in the particle number densities, and we employed the standard \KB ansatz for
their derivation. Further studies are needed in order to estimate whether and
in how far the situation for Boltzmann equations can be improved by including
non-minimal collision terms or by employing a generalized \KB ansatz.

In the future it will be important to perform a similar comparison of
Boltzmann and \KB equations also in the framework of gauge theories. Above
that a complete quantum mechanical description of leptogenesis would require
a treatment of \KB equations on an expanding space-time, which induces further
non-equilibrium effects. Independent of the comparison of Boltzmann and \KB
equations we are looking forward to learn to which extend an entirely
non-perturbative renormalization procedure affects the results quantitatively.
Above all, such a non-perturbative renormalization procedure should have a
stabilizing virtue for the computational algorithms.

\begin{acknowledgments}
  $M^3$ would like to thank J\"urgen Berges for collaboration on related work.
  Furthermore, we would like to thank Mathias Garny, Patrick Huber and Andreas
  Hohenegger for discussions and valuable hints, and Frank K\"ock for his
  continuous assistance with the computer cluster at our institute. Especially
  the early stages of this work were supported by the Technical University of
  Munich, the Max-Planck-Institute for Physics (Werner-Heisenberg-Institute)
  in Munich and the ``Sonderforschungsbereich 375 f\"ur Astroteilchenphysik
  der Deutschen Forschungsgemeinschaft''.
\end{acknowledgments}

\appendix

\section{Self Energies \label{sect3.2}}

In this appendix we give the expressions for the self energies, which have to
be inserted in the simplified \KB equations (\ref{eq3.38}) to (\ref{eq3.43}).
According to Eq.~(\ref{eq3.44}) the effective mass in Eqs.~(\ref{eq3.38}) and
(\ref{eq3.39}) is given by
\[ M^2 \left( x^0 \right) = m_B^2 + 24 \lambda \int \frac{\ddd{p}}{\left( 2 \pi \right)^3} G_F \left( x^0, x^0, p \right) \;. \]
Using the notation
\[ \bm{S_V} \left( x^0, y^0, \bm{k} \right) = \frac{\bm{k}}{k} S_V \left( x^0, y^0, k \right) \]
the statistical and spectral Higgs self-energies can be written in the form
\begin{widetext}
\begin{eqnarray*}
        \lefteqn{\Pi_F \left( x^0, y^0, k \right) = - 8 \eta^2 \int \frac{\ddd{p}}{\left( 2 \pi \right)^3} \int \ddd{q} \; \delta \left( \bm{k} - \bm{p} - \bm{q} \right)} \; \\
  &   & {} \times \Big[ - S^0_{V, F} \left( x^0, y^0, q \right) S^0_{V, F} \left( x^0, y^0, p \right) + \frac{1}{4} S^0_{V, \varrho} \left( x^0, y^0, q \right) S^0_{V, \varrho} \left( x^0, y^0, p \right) \\
  &   & {} + \bm{S_{V, F}} \left( x^0, y^0, \bm{q} \right) \bm{S_{V, F}} \left( x^0, y^0, \bm{p} \right) - \frac{1}{4} \bm{S_{V, \varrho}} \left( x^0, y^0, \bm{q} \right) \bm{S_{V, \varrho}} \left( x^0, y^0, \bm{p} \right) \Big]
\end{eqnarray*}
and
\begin{eqnarray*}
        \lefteqn{\Pi_{\varrho} \left( x^0, y^0, k \right) = - 16 \eta^2 \int \frac{\ddd{p}}{\left( 2 \pi \right)^3} \int \ddd{q} \; \delta \left( \bm{k} - \bm{p} - \bm{q} \right)} \; \\
  &   & {} \times \Big[ - S^0_{V, \varrho} \left( x^0, y^0, q \right) S^0_{V, F} \left( x^0, y^0, p \right) + \bm{S_{V, \varrho}} \left( x^0, y^0, \bm{q} \right) \bm{S_{V, F}} \left( x^0, y^0, \bm{p} \right) \Big] \;.
\end{eqnarray*}
The simplified lepton self-energies are given by
\begin{eqnarray*}
        \lefteqn{\Sigma_{V, F}^0 \left( x^0, y^0, k \right) = - 4 \eta^2 \int \frac{\ddd{p}}{\left( 2 \pi \right)^3} \int \ddd{q} \; \delta \left( \bm{k} - \bm{p} - \bm{q} \right)} \; \\
  &   & {} \times \Big[ G_F \left( x^0, y^0, q \right) S_{V, F}^0 \left( x^0, y^0, p \right) - \frac{1}{4} G_{\varrho} \left( x^0, y^0, q \right) S_{V, \varrho}^0 \left( x^0, y^0, p \right) \Big] \;,
\end{eqnarray*}
\begin{eqnarray*}
        \lefteqn{\Sigma_{V, \varrho}^0 \left( x^0, y^0, k \right) = - 4 \eta^2 \int \frac{\ddd{p}}{\left( 2 \pi \right)^3} \int \ddd{q} \; \delta \left( \bm{k} - \bm{p} - \bm{q} \right)} \; \\
  &   & {} \times \Big[ G_{\varrho} \left( x^0, y^0, q \right) S_{V, F}^0 \left( x^0, y^0, p \right) + G_F \left( x^0, y^0, q \right) S_{V, \varrho}^0 \left( x^0, y^0, p \right) \Big] \;,
\end{eqnarray*}
\begin{eqnarray*}
        \lefteqn{\Sigma_{V, F} \left( x^0, y^0, k \right) = - 4 \eta^2 \frac{\bm{k}}{k} \int \frac{\ddd{p}}{\left( 2 \pi \right)^3} \int \ddd{q} \; \delta \left( \bm{k} - \bm{p} - \bm{q} \right)} \; \\
  &   & {} \times \Big[ G_F \left( x^0, y^0, q \right) \bm{S_{V, F}} \left( x^0, y^0, \bm{p} \right) - \frac{1}{4} G_{\varrho} \left( x^0, y^0, q \right) \bm{S_{V, \varrho}} \left( x^0, y^0, \bm{p} \right) \Big] \;,
\end{eqnarray*}
and
\begin{eqnarray*}
        \lefteqn{\Sigma_{V, \varrho} \left( x^0, y^0, k \right) = - 4 \eta^2 \frac{\bm{k}}{k} \int \frac{\ddd{p}}{\left( 2 \pi \right)^3} \int \ddd{q} \; \delta \left( \bm{k} - \bm{p} - \bm{q} \right)} \; \\
  &   & {} \times \Big[ G_{\varrho} \left( x^0, y^0, q \right) \bm{S_{V, F}} \left( x^0, y^0, \bm{p} \right) + G_F \left( x^0, y^0, q \right) \bm{S_{V, \varrho}} \left( x^0, y^0, \bm{p} \right) \Big] \;.
\end{eqnarray*}
\end{widetext}

\section{Quantum-Kinetic Equations \label{sect3.3}}

Here we give the complete set of quantum kinetic equations, which are obtained
from the simplified \KB equations (\ref{eq3.38}) to (\ref{eq3.43}) once one
performs a Wigner transformation and a first-order gradient expansion. The
quantum-kinetic equations for the scalars read:
\begin{eqnarray}
        - \PB{\Omega}{G_F}
  & = & \Pi_{\varrho} G_F  - \Pi_F G_{\varrho} + \PB{\Pi_F}{\Re \left( G_R \right)} \;, \nonumber \\[5mm]
        - \PB{\Omega}{G_{\varrho}}
  & = & \PB{\Pi_{\varrho}}{\Re \left( G_R \right)} \;, \label{eq3.36} \\[5mm]
        G_R 
  & = & \frac{1}{- \omega^2 + k^2 + M^2 \left( t \right) + \Pi_R} \;. \nonumber
\end{eqnarray}
It can be shown that~\cite{Blaizot:2001nr}
\[ G_{\varrho} = 2 \; \Im \left( G_R \right) = \frac{- \Pi_{\varrho}}{\Omega^2 + \frac{1}{4} \Pi^2_{\varrho}} \]
indeed satisfies the kinetic equation for the scalar spectral function 
(\ref{eq3.36}). The quantum-kinetic equations for the fermionic statistical
propagators and spectral functions read:
\begin{eqnarray*}
  \lefteqn{\PB{W}{S_{V, F}^0} = \Sigma_{V, \varrho}^0 S_{V, F}^0 - \Sigma_{V, F}^0 S_{V, \varrho}^0 + \Sigma_{V, \varrho} S_{V, F}} \; \\
  &   & {} - \Sigma_{V, F} S_{V, \varrho} - \PB{\Sigma_{V, F}^0}{\Re \left( S_{V, R}^0 \right)} \\
  &   & {} + \PB{\Re \left( \Sigma_{V, R} \right)}{S_{V, F}} + \PB{\Sigma_{V, F}}{\Re \left( S_{V, R} \right)} \;,
\end{eqnarray*}
\begin{eqnarray*}
  \lefteqn{\PB{W}{S_{V, F}} = \Sigma_{V, \varrho}^0 S_{V, F} + \Sigma_{V, F}^0 S_{V, \varrho} + \Sigma_{V, \varrho} S_{V, F}^0} \; \\
  &   & {} + \Sigma_{V, F} S_{V, \varrho}^0 + \PB{\Re \left( \Sigma_{V, R} \right)}{S_{V, F}^0} \\
  &   & {} - \PB{\Sigma_{V, F}^0}{\Re \left( S_{V, R} \right)} + \PB{\Sigma_{V, F}}{\Re \left( S_{V, R}^0 \right)} \;,
\end{eqnarray*}
\begin{eqnarray*}
  \lefteqn{\PB{W}{S_{V, \varrho}^0} = - \PB{\Sigma_{V, \varrho}^0}{\Re \left( S_{V, R}^0 \right)}} \; \\
  &   & {} - \PB{\Re \left( \Sigma_{V, R} \right)}{S_{V, \varrho}} - \PB{\Sigma_{V, \varrho}}{\Re \left( S_{V, R} \right)} \;,
\end{eqnarray*}
and
\begin{eqnarray*}
  \lefteqn{\PB{W}{S_{V, \varrho}} = \PB{\Sigma_{V, \varrho}^0}{\Re \left( S_{V, R} \right)}} \; \\
  &   & {} - \PB{\Re \left( \Sigma_{V, R} \right)}{S_{V, \varrho}^0} + \PB{\Sigma_{V, \varrho}}{\Re \left( S_{V, R}^0 \right)} \;.
\end{eqnarray*}
The quantum-kinetic equations for the retarded lepton propagators read
\[ S_{V, R}^0 = \frac{W - \frac{i}{2} \Sigma_{V, \varrho}^0}{\left( W - \frac{i}{2} \Sigma_{V, \varrho}^0 \right)^2 - \left( k - \Re \left( \Sigma_{V, R} \right) - \frac{i}{2} \Sigma_{V, \varrho} \right)^2} \]
and
\[ S_{V, R} = - \frac{k - \Re \left( \Sigma_{V, R} \right) - \frac{i}{2} \Sigma_{V, \varrho}}{\left( W - \frac{i}{2} \Sigma_{V, \varrho}^0 \right)^2 - \left( k - \Re \left( \Sigma_{V, R} \right) - \frac{i}{2} \Sigma_{V, \varrho} \right)^2} \;. \]

\section{Simplifying the Boltzmann Collision Integrals \label{sect3.4}}

This appendix reveals the details of the calculation leading from the
Boltzmann equations (\ref{eq3.24}) and (\ref{eq3.25}) to their simplified
versions (\ref{eq3.29}) and
(\ref{eq3.30})~\cite{Lindner:2005kv,Dolgov:1997mb}. For zero momentum the
evaluation of the collision integral in Eq.~(\ref{eq3.24}) is literally
trivial:
\begin{widetext}
\begin{eqnarray*}
        \partial_t n_s \left( t, k = 0 \right)
  & = & - \frac{m \eta^2}{4 \pi} \bigg[ \Big( n_s \left( t, k = 0 \right) + 1 \Big) n_f \left( t, p \right) n_f \left( t, q \right) \\
  &   & \qquad {} - n_s \left( t, k = 0 \right) \Big( n_f \left( t, p \right) - 1 \Big) \Big( n_f \left( t, q \right) - 1 \Big) \bigg]_{p = q = \frac{m}{2}} \;.
\end{eqnarray*}
\end{widetext}
For $k>0$ a little more work has to be done. We rewrite Eq.~(\ref{eq3.24}) 
using the Fourier representation of the momentum conservation $\delta$ 
function
\[ \delta^3 \left( \bm{k} - \bm{p} - \bm{q} \right) = \int \frac{\ddd{\xi}}{\left( 2 \pi \right)^3} \;\exp \left( - i \bm{k} \bm{\xi} + i \bm{p} \bm{\xi} + i \bm{q} \bm{\xi} \right) \]
and spherical coordinates. The scalar product of two vectors is then given by
\[ \bm{p} \bm{q} = p q \Big( \sin \vartheta_p \; \sin \vartheta_q \; \cos \left( \varphi_p - \varphi_q \right) + \cos \vartheta_p \; \cos \vartheta_q \Big) \;. \]
We perform the integrals over the solid angles in the order $\Omega_q$,
$\Omega_p$, $\Omega_{\xi}$. Using the notation
\[ j \left( x \right) = \frac{\sin \left( x \right)}{x} - \cos \left( x \right) \]
we find
\begin{widetext}
\begin{eqnarray*} 
        \int d\Omega_q \; \exp \left( i \bm{q} \bm{\xi} \right) \left( \frac{\bm{p} \bm{q}}{pq} - 1 \right)
  & = & \frac{4 \pi}{q \xi} \Big( i \cos \left( \vartheta_p \right) j \left( q \xi \right) - \sin \left( q \xi \right) \Big) \;, \\
        \int d\Omega_p \; \exp \left( i \bm{p} \bm{\xi} \right) \Big( i \cos \left( \vartheta_p \right) j \left( q \xi \right) - \sin \left( q \xi \right) \Big)
  & = & - \frac{4 \pi}{p \xi} \Big( j \left( p \xi \right) j \left( q \xi \right) + \sin \left( p \xi \right) \sin \left( q \xi \right) \Big) \;, \\
        \int d\Omega_{\xi} \exp \left( - i \bm{k} \bm{\xi} \right)
  & = & \frac{4 \pi}{k \xi} \sin \left( k \xi \right) \;.
\end{eqnarray*}
After defining the auxiliary function
\begin{eqnarray*}
        J_s \left( k, p, q \right)
  & = & - p q \intl_0^{\infty} d\xi \; \frac{\sin \left( k \xi \right)}{k \xi} \Big( j \left( p \xi \right) j \left( q \xi \right) + \sin \left( p \xi \right) \sin \left( q \xi \right) \Big) \\
  & = & - \frac{\pi}{16 k} \Big( k^2 - \left( p + q \right)^2 \Big) \Big( \sign \left( k - p - q \right) - \sign \left( k + p - q \right) \\
  &   & {} - \sign \left( k - p + q \right) + \sign \left( k + p + q \right) \Big)
\end{eqnarray*}
\end{widetext}
and integrating over $q$, we eventually arrive at Eq.~(\ref{eq3.29}), where 
\[ q_0 = E \left( k \right) - p \;. \]
Next, we work out the collision integral for the fermions. First of all, we 
integrate Eq.~(\ref{eq3.25}) over $\Omega_k$. On the left hand side this gives
a factor of $4 \pi$. On the right hand side we evaluate the integrals over the
solid angles in the order $\Omega_q$, $\Omega_p$, $\Omega_k$, $\Omega_{\xi}$:
\begin{widetext}
\begin{eqnarray*}
        \int d\Omega_q \; \exp \left( i \bm{q} \bm{\xi} \right)
  & = & \frac{4 \pi}{q \xi} \sin \left( q \xi \right) \;, \\
        \int d\Omega_p \; \exp \left( - i \bm{p} \bm{\xi} \right) \left( \frac{\bm{k} \bm{p}}{k p} + 1 \right)
  & = & \frac{4 \pi}{p \xi} \Big( \sin \left( p \xi \right) - i \cos \left( \vartheta_k \right) j \left( p \xi \right) \Big) \;, \\
        \int d\Omega_k \; \exp \left( - i \bm{k} \bm{\xi} \right) \left( \sin \left( p \xi \right) - i \cos \left( \vartheta_k \right) j \left( p \xi \right) \right)
  & = & \frac{4 \pi}{k \xi} \Big( \sin \left( k \xi \right) \sin \left( p \xi \right) - j \left( k \xi \right) j \left( p \xi \right) \Big) \;, \\
        \int d\Omega_{\xi}
  & = & 4 \pi \;.
\end{eqnarray*}
Defining the auxiliary function
\begin{eqnarray*}
        J_f \left( k, p, q \right)
  & = & p \intl_0^{\infty} d\xi \; \frac{\sin \left( q \xi \right)}{k \xi} \Big( \sin \left( k \xi \right) \sin \left( p \xi \right) - j \left( k \xi \right) j \left( p \xi \right) \Big) \\
  & = & \frac{\pi}{16 k^2} \Big( \left( k - p \right)^2 - q^2 \Big) \Big( \sign \left( k - p - q \right) - \sign \left( k + p - q \right) \\
  &   & {} - \sign \left( k - p + q \right) + \sign \left( k + p + q \right) \Big)
\end{eqnarray*}
\end{widetext}
and integrating over $p$ yields Eq.~(\ref{eq3.30}), where 
\[ p_0 = E \left( q \right) - k \;. \]


\end{document}